\documentclass[a4paper]{jpconf}

\usepackage{graphicx}
\usepackage[caption=false]{subfig}
\usepackage{amsmath} 
\usepackage{mathtools}
\usepackage{bm}
\usepackage{float}
\usepackage{epstopdf}
\usepackage{color}
\usepackage{enumerate}

\newcommand{\beq}{\begin{equation}}
\newcommand{\enq}{\end{equation}}
\newcommand{\beqn}{\begin{eqnarray}}
\newcommand{\enqn}{\end{eqnarray}}

\newcommand{\al}{\alpha}
\newcommand{\be}{\beta}
\newcommand{\ga}{\gamma}
\newcommand{\de}{\delta}
\newcommand{\ep}{\epsilon}

\newcommand{\f}[2]{\frac{#1}{#2}}

\newcommand{\h}[1]{\hat{#1}}
\newcommand{\dg}{\dagger}

\newcommand{\bra}[1]{\langle #1|}
\newcommand{\ket }[1]{|#1\rangle}

\newcommand{\mU}{\mathcal{U}}
\newcommand{\mV}{\mathcal{V}}

\newcommand{\ii}{\textrm{i}}

\begin{document}
\title{Many-body Propagator Theory with Three-Body Interactions: a Path to Exotic Open Shell Isotopes}

\author{C. Barbieri}

\address{Department of Physics, University of Surrey, Guildford GU2 7XH, UK}

\ead{C.Barbieri@surrey.ac.uk}

\begin{abstract}
 Ab-initio predictions of nuclei with masses up to A$\sim$100 or more are becoming possible thanks to novel advances in computations and in the formalism of many-body physics. Some of the most fundamental issues include how to deal with many-nucleon interactions, how to calculate degenerate---open shell---systems, and pursuing ab-initio approaches to reaction theory. Self-consistent Green's function (SCGF) theory is a natural approach to address these challenges.  Its formalism has recently been extended to three- and many-body interactions and reformulated within the Gorkov framework to reach semi-magic open shell isotopes. These exciting developments, together with the predictive power of chiral nuclear Hamiltonians, are opening the path to understanding large portions of the nuclear chart, especially within the {\em sd} and {\em pf} shells. The present talk reviews the most recent advances in ab-initio nuclear structure and many-body theory that have been possible through the SCGF approach. 
 
\end{abstract}

\section{Introduction} ~ \\
Microscopic first principle predictions of atomic nuclei are highly desirable since they can impact research of exotic isotopes.
They could help in constraining extrapolations to higher mass regions~\cite{Erler2012nature} and to extreme proton-neutron asymmetries~\cite{Waldecker:2011by}, including regions close to the drip lines where experimental data will be unavailable for the foreseeable future.
Such an ambitious program requires advances in quantum many-body theory that include extending existing methods to deal with many-particle interactions, developing new approaches to calculate degenerate (open shell) fermionic systems, and calculating microscopic optical potentials.

Ab-initio methods such as coupled-cluster (CC)~\cite{Hagen2010abinit,Binder:2012mk}, in-medium similarity renormalization group (IMSRG)~\cite{Tsukiyama2011prl,Hergert2013a} or self-consistent Green's function (SCGF) theory based on the Dyson equation~\cite{Dickhoff2004ppnp,Barbieri:2007Atoms,Barbieri2009ni} (Dyson-GF) have accessed medium-mass nuclei up to A$\sim$56 on the basis of realistic two-nucleon (NN) interactions.
However, it has become clear that three-nucleon forces (3NFs) play a major role in determining crucial features of exotic isotopes, such as the evolution of magic numbers and the position of driplines~\cite{Otsuka2010prl,Holt2011jpg,Hagen2012prlOx,Hergert2013prl,Cipollone2013prl}. Realistic NN and three-nucleon (3N) interactions based on chiral perturbation theory have been used in Lattice Effective Field Theory~\cite{Lahde2014plb,Epelbaum2014prl} but also  evolved to low momentum cutoffs, retaining both induced and pre-existing 3NFs~\cite{Jurgenson2009prl,Roth2012prl}. Systematic implementations of similar Hamiltonians within the above many-body methods are required to eventually achieve quantitative predictions of observables for medium-mass isotopes. This has been addressed only recently for  CC~\cite{Hagen2007cc3nf,Binder2013cc3nf} and SCGF~\cite{Carbone2013tnf} methods. Based on these developments, closed shell nuclei up to Sn can now be approached~\cite{Binder2014ccSn132}.

A second and major challenge to ab-initio theory is that standard implementations of the above methods are  limited to doubly closed (sub-)shell nuclei and to their immediate neighbors~\cite{Jansen:2011cc,Cipollone2013prl}. As one increases the nuclear mass, longer chains of truly open shell nuclei emerge that cannot be accessed with these approaches. Many-body techniques that could tackle genuine open shell systems---or, at least, singly open shells---would immediately extend the reach of ab-initio studies from a few tens to several hundreds of mid-mass isotopes.  Our collaboration has proposed to exploit ideas based on breaking particle-number symmetry in order to achieve this goal~\cite{ParisProc1,ParisProc2,Soma:2011GkvI}.  
This has led to reformulating the SCGF in the Gorkov formalism (Gorkov-GF)~\cite{Soma:2011GkvI,Soma:2013rc,Soma2014GkvII}, which will be discussed in this talk. Applications of Gorkov-GF have been successful for the Ca and neighbouring isotopes up to $^{54}$Ti~\cite{Soma2014s2n}.
Recently, similar developments have also been introduced within the IMSRG framework~\cite{Hergert2013prl}.

The last challenge to  theory is to provide consistent descriptions of the structure and reactions of nuclei in order to constrain and improve the analysis of experimental data. Microscopic calculations of elastic nucleon scattering were achieved, e.g., for a $^{16}$O target in~\cite{Barbieri:2005NAscatt,Hagen2010F17scattt}. Recently, ab-initio calculations, including cluster projectiles,  have been possible for few-body targets by combining the ab-initio no-core shell model (NCSM)  with the resonating-group method (RGM)~\cite{Quaglioni2012prl,Baroni2013prl}. For larger masses, the SCGF becomes the method of choice due to the equivalence between the many-body self-energy and Fesbach theory of elastic scattering. This fact has guided recent advances of phenomenological dispersive optical models (DOM)~\cite{Charity2006prl,Mahzoon2014prl} and microscopic calculations of optical potentials~\cite{Waldecker:2011by,Barbieri:2005NAscatt,Polls2011dom}. Further developing these approaches to proper ab-initio methods will be crucial to advance our understanding of exotic radioactive beam experiments.

This talk reports about recent progress on the above topics within the SCGF approach to quantum many-body physics.
The new details of the formalism are discussed first and results for finite nuclei are reported independently in sections~\ref{SecOptMod} and~\ref{Sec3NF}.

\section{Propagator theory} 
~ \\
In Green's function (or propagator) theory one calculates the single particle propagator~\cite{Fett1971book}:
\begin{equation}
 g_{\alpha \beta}(\omega) ~=~ 
 \sum_n  \frac{ 
          \langle {\Psi^A_0}     \vert c_\alpha        \vert {\Psi^{A+1}_n} \rangle
          \langle {\Psi^{A+1}_n} \vert c^{\dag}_\beta  \vert {\Psi^A_0} \rangle
              }{\omega - (E^{A+1}_n - E^A_0) + i \eta }  ~+~
 \sum_k \frac{
          \langle {\Psi^A_0}     \vert c^{\dag}_\beta  \vert {\Psi^{A-1}_k} \rangle
          \langle {\Psi^{A-1}_k} \vert c_\alpha        \vert {\Psi^A_0} \rangle
             }{\omega - (E^A_0 - E^{A-1}_k) - i \eta } \; ,
\label{eq:g1}
\end{equation}
where $\alpha$,$\beta$,..., label a complete orthonormal basis set. 
In Eq.~\eqref{eq:g1}, $\vert\Psi^{A+1}_n\rangle$, $\vert\Psi^{A-1}_k\rangle$ 
are the eigenstates, and $E^{A+1}_n$, $E^{A-1}_k$ the eigenenergies of the 
($A\pm1$)-nucleon system. Therefore, the poles of the propagator reflect experimental
transfer energies for the addition or the separation of a nucleon to and from the ground
state of the A-body nucleus.
We obtain this information by solving the Dyson equation,
\begin{equation}
\label{eq:Dyson}
g_{\alpha \beta}(\omega) = g_{\alpha \beta}^{\text{HF}}(\omega) + \sum_{\gamma \, \delta} \,
g_{\alpha \gamma}^{\text{HF}}(\omega) \, \Sigma^{\star}_{\gamma \delta}(\omega) \, g_{\delta \beta}(\omega) \, ,
\end{equation}
which recasts the many-body Schr\"odinger equation into an equation for the {\em dressed} (i.e. fully correlated) propagator $g(\omega)$.
In Eq.~\eqref{eq:Dyson}, $g^{\text{HF}}(\omega)$ represents the unperturbed propagator, which is taken to be an Hartree-Fock (HF) state for all applications discussed in the following.
The information on the many-body dynamics is coded in the irreducible self-energy $\Sigma^{\star}(\omega)$. This has a Lehman representation similar to Eq.~\eqref{eq:g1}:
\begin{equation}
\label{eq:selfenergy}
\Sigma^{\star}_{\alpha \beta}(\omega) = \Sigma^{{\star},(\infty)}_{\alpha \beta} + 
\sum_{n \, n'} C_{\alpha n} \left [ \frac{1}{\omega-M} \right ]_{n n'} C^{\dagger}_{n' \beta} +
\sum_{p \, p'} D_{\alpha p} \left [ \frac{1}{\omega-N} \right ]_{p p'} D^{\dagger}_{p' \beta} 
\end{equation}
and provides the microscopic optical potential for a single nucleon interacting with the whole many-body system.
In Eq.~\eqref{eq:selfenergy}, $M, N$ are interaction matrices in the $2h1p$ and $2p1h$ (or more complex) configuration spaces and $C, D$ are the respective couplings to single-particle states. $\Sigma^{{\star},(\infty)}$ represents the static (i.e., energy independent) contribution to the self-energy.

\begin{figure}
\begin{center}
\includegraphics[height=.20\textheight]{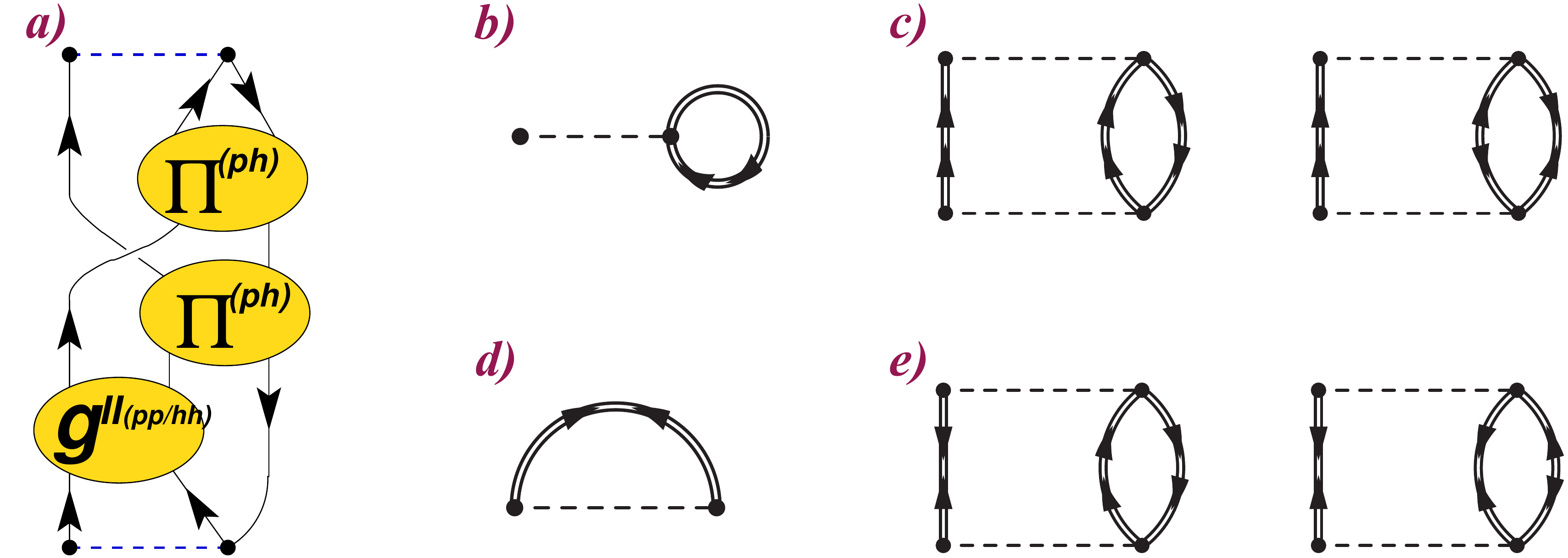}
\end{center}
\vskip  -.3cm
  \caption{  \small {\em Left}: Diagram a) gives one example of the particle-vibration coupling terms that enter the ADC(3) and FRPA expansions of the self-energy. 
 {\em Center}~and {\em right}: Contribution to the Gorkov irreducible self-energy up to second order, for the case of NN interactions only.
 Diagram b) is the first order contributions to the normal part of the static irreducible self-energy, $\Sigma^{11, (\infty)}$. 
 This also appears in the standard Dyson formulation and extends the usual HF potential to the one corresponding to a fully correlated density matrix.
 Diagram d) is the anomalous first order contribution $\Sigma^{21, (\infty)}$, which corresponds to a correlated version of the pairing potential in the HFB equations.
 Diagrams c) are the normal second order terms and diagrams e) are the anomalous second order ones.
}
\label{fig:se_exp}
\end{figure}

The self-energy is constructed starting from the Hamiltonian $H(A) = H - \hat{T}_{\text{c.o.m.}}(A) = \hat{U}(A) + \hat{V}(A) + \hat{W}$, where we correct for the centre-of-mass kinetic energy. $\hat{U}$, $\hat{V}$ and $\hat{W}$ collect all one-, two- and three-body interactions, respectively. 
%
%
For finite closed-shell nuclei, the propagator~\eqref{eq:g1} is calculated
by first solving spherical HF equations.
The $g^{\text{HF}}(\omega)$ propagator is then used as a reference state to calculate the self-energy, using either
the third-order algebraic diagrammatic construction [ADC(3)]~\cite{Schirmer1982ADC2,Schirmer1983ADCn} or
the Faddeev random phase approximation (FRPA)~\cite{Barbieri:2001frpa,Barbieri:2007Atoms}
methods. Both the ADC(3) and the FRPA completely include all diagrams up to third-order
and resum many others to all orders.
Moreover, they completely accounts for particle-vibration diagrams as shown in Fig.~\ref{fig:se_exp}a. 
Thus, in general, a full FRPA calculation involves the calculations of the particle-hole ($ph$)  response
and of the two-body spectral function. The latter has been employed, for example, to investigate nuclear
correlations from two-nucleon emission experiments~\cite{Barbieri2004Shh,Middleton:2006O16pn,Middleton2010gpN}.
The difference between the two approaches is in the fact that vibrations in the $ph$
and particle-particle/hole-hole ($pp$/$hh$) channels are calculated in Tamn-Dancoff approximation
for ADC(3) and in random phase approximation for FRPA.

It is instructive to compare the perturbation theory content of the ADC($n$) and FRPA methods to
that of other many-body approaches. For the calculation of total energies, both ADC(3)/FRPA
and the coupled cluster CCSD approaches retain in full all diagrams up to second order~\cite{Trofimov2005nDadc}.
However, the Koltun energy sum rule  in SCGF theory adds specific triple corrections.  Our experience
with both G-matrix and SRG-evolved interactions is that ADC(3) results are indeed systematically closer
to CCSD(T)  rather than to the simple CCSD~\cite{Soma:2013rc}.
When evaluating  one nucleon addition and separation energies, the ADC(3) fully sums
two-particle--one-hole ($2p1h$) and two-hole--one-particle ($2h1p$) configurations and should
therefore be compared to the usual truncations of the excitation operator in particle-attached 
and particle-removed CC. This is usually sufficient to predict the dominant quasiparticle and quasihole
fragments of the spectral distribution.
Other small fragments---the so-called shakeup states---will generally require truncation
schemes beyond $2p1h$/$2h1p$, such as the ADC(4) and ADC(5)~\cite{Schirmer1983ADCn}.

\section{Extension of propagator theory to include three-body interactions}
\label{sec:3nf}
\begin{figure}
 \includegraphics[width=0.60\columnwidth]{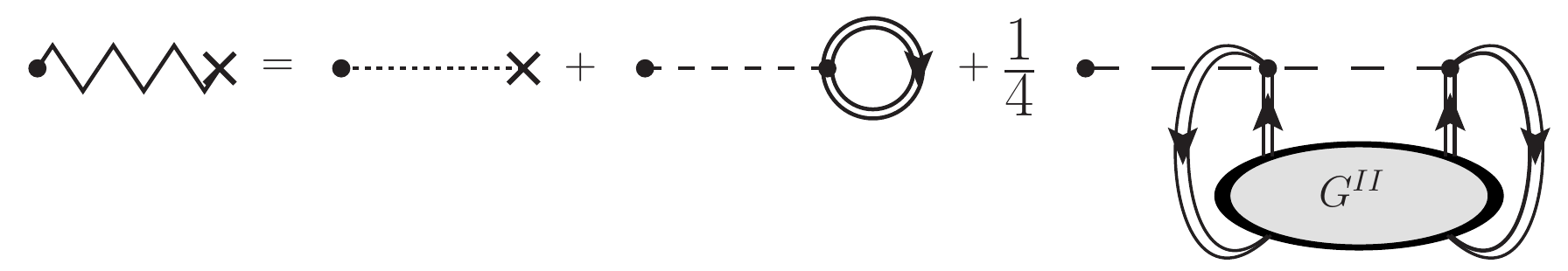}
\\   \\
\includegraphics[width=0.45\columnwidth]{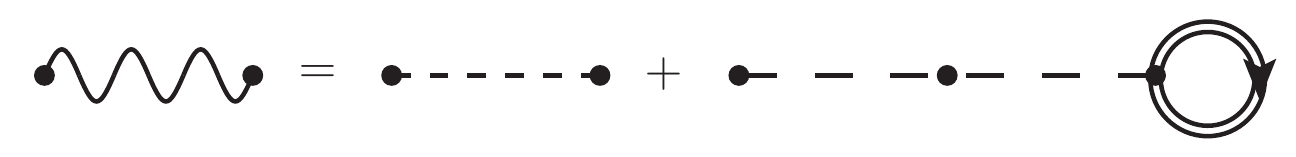}
\begin{center}
\caption{ \small {\em Top:} Diagrammatic representation of the effective one-body interaction of Eq.~(\ref{ueff}). This is given by the sum of the original one-body potential $\hat{U}$ (dotted line), the two-body interaction $\hat{V}$ (dashed line) contracted with a dressed one-body propagator $g$ (double line with arrow), and the three-body interaction $\hat{W}$ (long-dashed line) contracted with a dressed two-body propagator, $g^{II}$. The correct symmetry factor of 1/4 is also shown explicitly. 
{\em Bottom:} Diagrammatic representation of the effective two-body interaction of Eq.~(\ref{veff}). This is given by the sum 
of the original two-body interaction and the three-body interaction contracted with a dressed propagator.}
\label{fig:ueffective}
\end{center}
\end{figure}
~ \\
We have extended the Dyson SCGF formalism to the case of Hamiltonians which include three-body interactions.
Full details are presented in Refs.~\cite{Carbone2013tnf,Carbone2014phd} and involve defining in-medium effective interactions that regroup specific sets of diagrams, such that only {\em interaction-irreducible} ones are retained\footnote{A diagram is said to be {\em interaction-reducible} if it can be factorized in two lower order diagrams by cutting an interaction vertex or, equivalently, if it is connected and there exists a group of lines (whether interacting among themselves or not) that leave an interaction vertex and eventually all return to it.}.
 Moreover, proper corrections to the Kotlun sum rule are required to calculate  total binding energies.
For a pure two-body Hamiltonian, the only possible interaction-reducible contribution to the self-energy is the generalised Hartree-Fock diagram of Fig.~\ref{fig:se_exp}b\footnote{For the Gorkov formalism, one would also have the interaction-reducible Bogoliubov potential of Fig.~\ref{fig:se_exp}d.}.
 However, many more appear when three- and  many-body forces are present. This makes it useful to define effective interactions to group such contributions. 

Hence, for a system with up to 3NFs, we define an effective Hamiltonian,
\begin{equation}
\widetilde H= {\widetilde U} + {\widetilde  V} + \h W \,
\label{Heff}
\end{equation}
where $\widetilde U$ and  $\widetilde V$ represent effective interaction operators. 
As long as only interaction-irreducible diagrams are calculated when using these effective interactions, no double counting of any diagram is possible~\cite{Carbone2013tnf}.
The explicit expressions for the one- and two-body effective interaction operators are:
\begin{eqnarray}
\label{ueff}
\widetilde U& =&\sum_{\al\be}\left[ U_{\al \be} 
-  i \hbar \sum_{\ga\de}V_{\al\ga,\be\de} \, g_{\de \ga}(t-t^+) 
+ \f{ i \hbar}{4} \sum_{\substack{\ga\ep \\ \de\eta}} W_{\al\ga\ep,\be\de\eta}
\,g^{II}_{\de\eta , \ga\ep}(t-t^+)\right] a_\al^\dg a_{\be}\,, \\ 
\label{veff}
\widetilde V &=& \f 1 4\sum_{\substack{\al\ga\\\be\de}}\left[V_{\al\ga,\be\de}
- i \hbar \sum_{\ep\eta}W_{\al\ga\ep,\be\de\eta} \,g_{\eta\ep}(t-t^+)\right] a_\al^\dg a_\ga^\dg a_{\de}a_{\be} \, ,
\end{eqnarray}
where we have introduced specific orientations of the one-body, $g(\tau)$, and two-body, $g^{II}(\tau)$, green
functions in time representation~\cite{Dickhoff:book}. Eqs.~\eqref{ueff} and~\eqref{veff} are shown
diagrammatically in Fig.~\ref{fig:ueffective}.

\begin{figure}[t]
\begin{center}
\includegraphics[width=0.83\columnwidth]{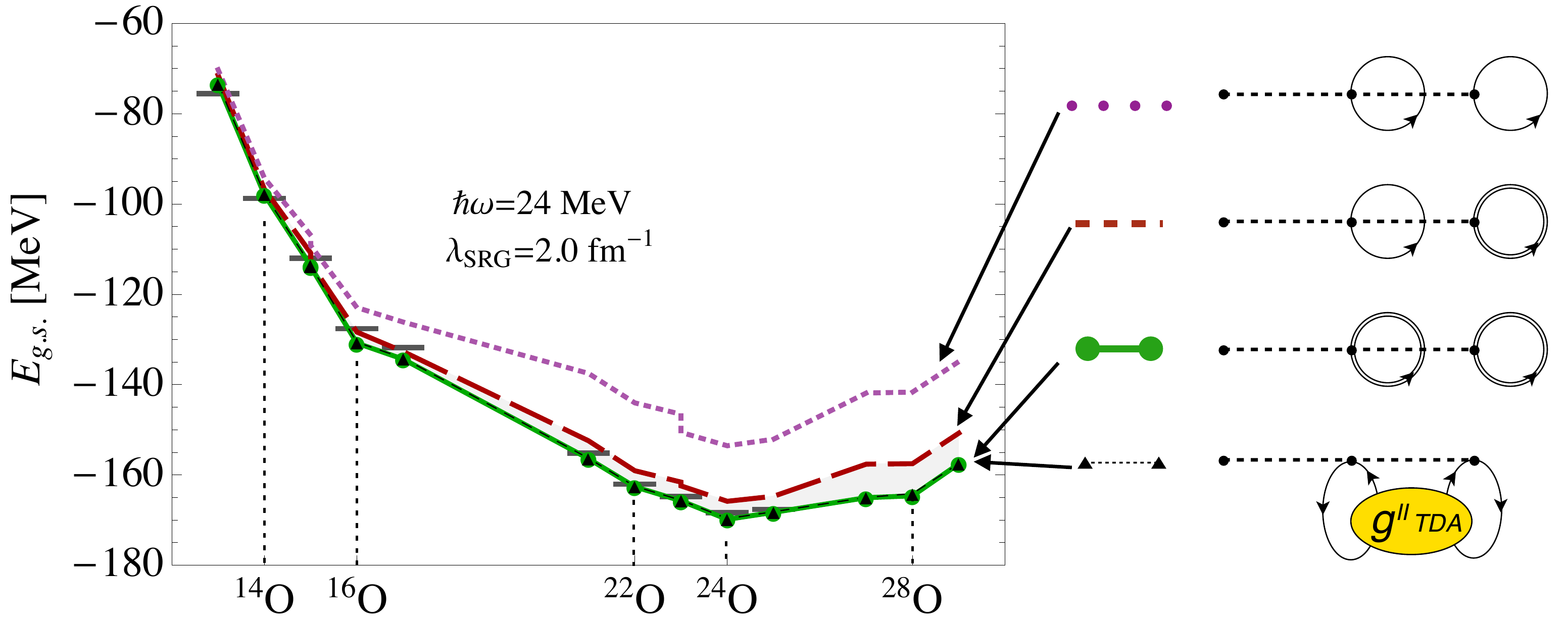}
\caption{ \small
 Calculated ground state energies of oxygen isotopes for different approximations of the 3NFs contributions to the static self-energy, see Eqs.~\eqref{ueff} and~\eqref{eq:1ord}. All results are obtained from a SRG evolved NN+3N interactions with cutoff $\lambda_{SRG}$=2 fm$^{-1}$ and are based on the full ADC(3) self-energy~\cite{Cipollone2013prl}. However, the last term in Eq.~\eqref{ueff} is calculated at different levels of self-consistency by contracting the 3NF with all unperturbed HF propagators (single line loops), by contracting with one dressed propagator (double line loops), or with two dressed propagators. The corresponding diagrams are shown on the right. Dressing self-consistently {\em all} the single-particle propagators is mandatory to obtain convergence with respect to the many-body truncation. Higher order contributions  from ladder diagrams (TDA ladders in the figure) do not further contribute for the present interaction.
 }
\label{fig:Ueff_conv}
\end{center}
\end{figure}

The use of effective interactions greatly reduces the number of diagrams to be accounted for in practical 
calculations. By construction, the only possible interaction-irreducible contribution at first-order
is precisely given by Eq.~\eqref{ueff}, and it equals the static irreducible self-energy:
\begin{equation}
\Sigma^{\star , (\infty)}_{\al \be} = \widetilde{U}_{\al \be} \; . 
\label{eq:1ord}
\end{equation}
Even though it only contributes to the self-energy at first order, the effective one-body potential $\widetilde{U}$
is very important. It represents the (energy-independent) mean field seen by each particle, including 
contributions from all many-body interactions.
The simplest approximation to $\widetilde{U}$ would consists in performing the averages of Eq.~\eqref{ueff}
over Hartree-Fock or other Slater determinant  reference states. This is equivalent to reducing $\widetilde{U}$
to the one-body term of a normal ordered Hamiltonian. Analogous treatments of the effective interaction
have been employed in nuclear physics calculations up to date, including both
finite nuclei~\cite{Otsuka2010prl,Roth2012prl,Hagen2012prlOx,Hagen2012prlCa}
 and nuclear matter~\cite{Soma:2008nn,Soma:2009pf,Hebeler:2011snm,Carbone2013snm} applications.
For the particular case of {\em ab-initio} SCGF calculations, we find that it is mandatory to fully dress the reference 
propagators entering Eq.~\eqref{ueff}~\cite{Cipollone2013prl}. This is shown in Fig.~\ref{fig:Ueff_conv},
where the last term of Eq.~\eqref{ueff}  is gradually improved.
While dressing the propagators that enter the HF contribution coming from 3NFs is mandatory, additional NN correlations
such as ladder series do not sensibly affect the final results%
\footnote{For methods like CC, such dressing of the mean field is  generated implicitly
when solving the CC equations iteratively. However, in the SCGF philosophy, this is included directly in the effective
interaction $\widetilde{U}$ and the dressed propagator can be thought of as an improved reference state.}.

\begin{figure}
  \centering
  \subfloat[]{\label{fig:2ord_2B}\includegraphics[scale=0.6]{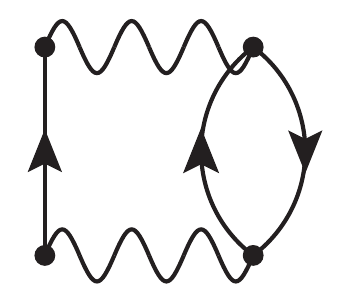}}
  \hspace{2cm}
  \subfloat[]{\label{fig:2ord_3B}\includegraphics[scale=0.6]{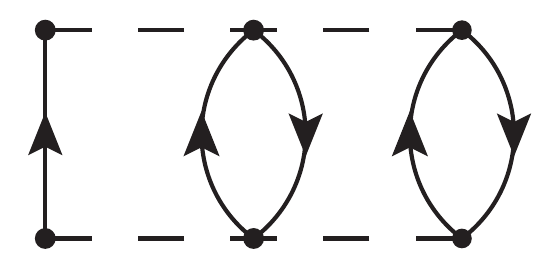}}
  \caption{ \small \emph{Interaction-irreducible}  diagrams contributing to the irreducible self-energy at  second order.
  These are the only contributions that must be considered when using the
   effective Hamiltonian of Eq.~(\ref{Heff}).}
  \label{fig:2ord}
\end{figure}

\begin{figure}[t]
  \centering
  \subfloat[]{\label{fig:3ord_2B_1}\includegraphics[scale=0.50]{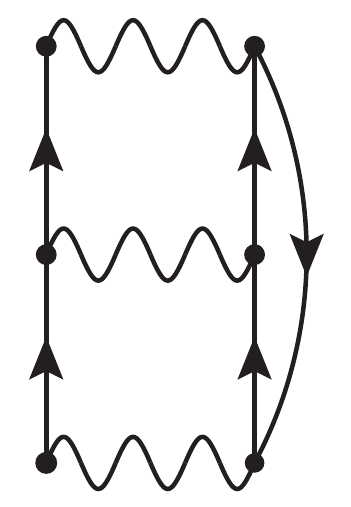}}
  \hspace{1.5cm}
  \subfloat[]{\label{fig:3ord_2B_2}\includegraphics[scale=0.50]{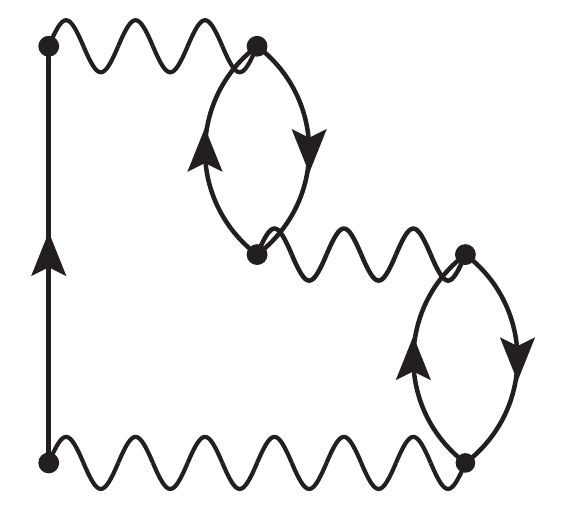}}
  \hspace{3.5cm}
  \subfloat[]{\label{fig:3ord_232B}\includegraphics[scale=0.50]{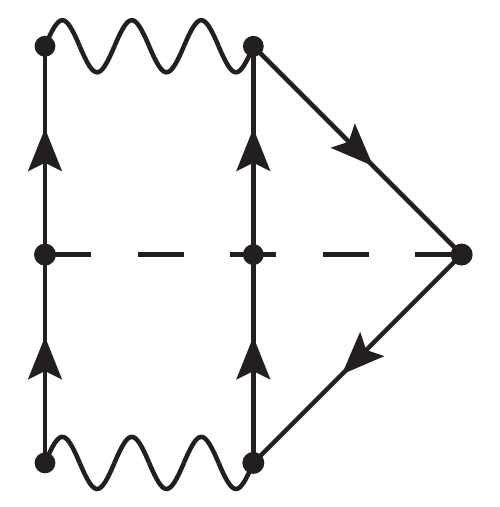}}
  \newline    \vskip .6cm
  \subfloat[]{\label{fig:3ord_223B_1}\includegraphics[scale=0.50]{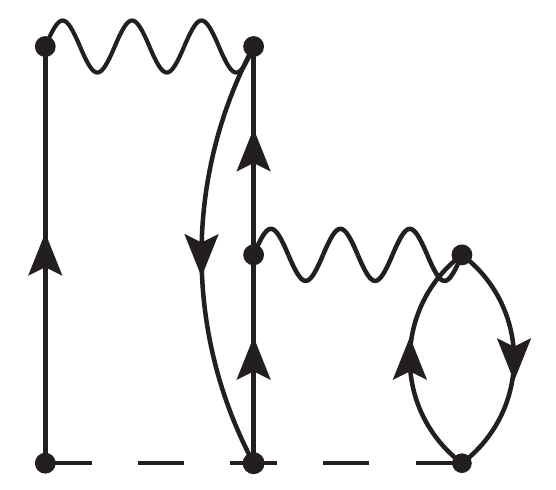}}
  \hspace{1.3cm}
  \subfloat[]{\label{fig:3ord_223B_2}\includegraphics[scale=0.50]{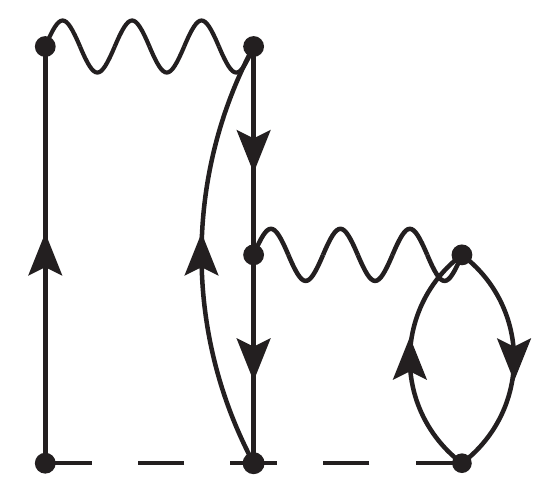}}
  \hspace{1.3cm}
   \subfloat[]{\label{fig:3ord_322B_1}\includegraphics[scale=0.50]{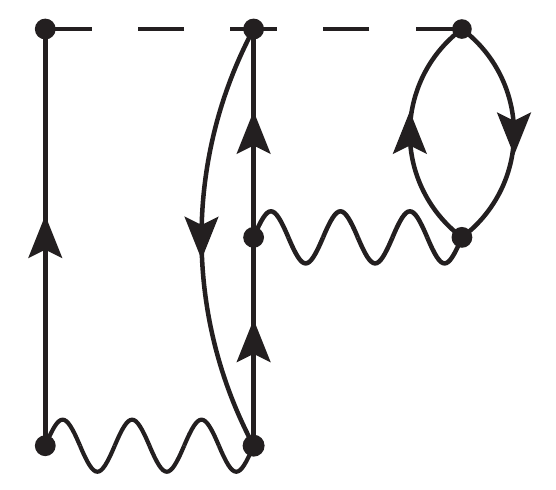}}
  \hspace{1.3cm}
  \subfloat[]{\label{fig:3ord_322B_2}\includegraphics[scale=0.50]{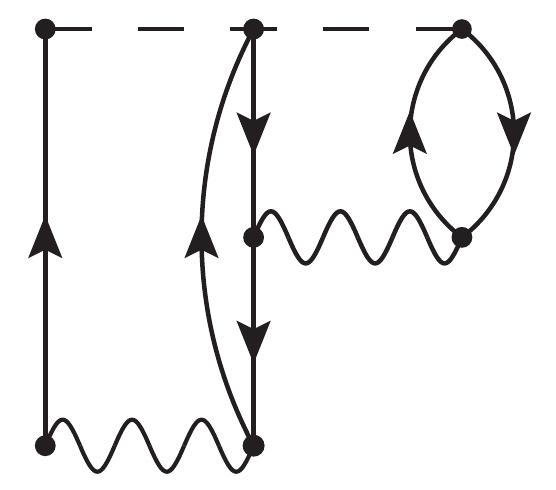}}
  \newline   \vskip .6cm
  \subfloat[]{\label{fig:3ord_233B_1}\includegraphics[scale=0.50]{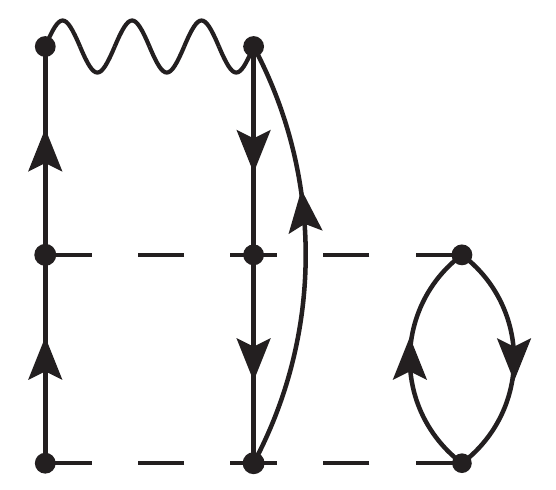}}
  \hspace{1.3cm}
  \subfloat[]{\label{fig:3ord_233B_2}\includegraphics[scale=0.50]{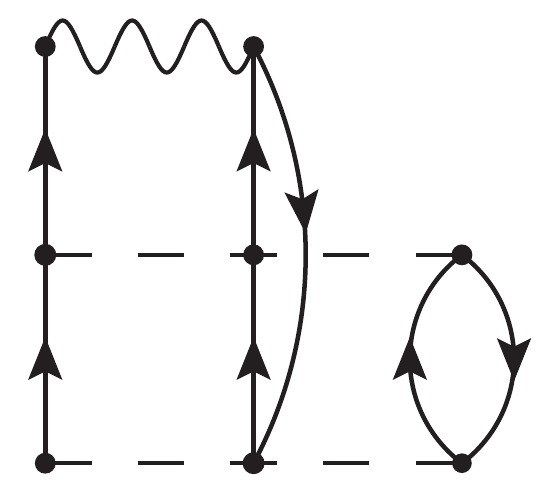}}
  \hspace{1.3cm}
  \subfloat[]{\label{fig:3ord_332B_1}\includegraphics[scale=0.50]{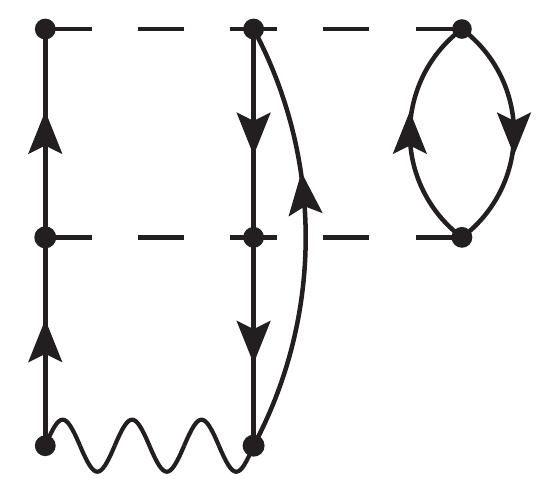}}
  \hspace{1.3cm}
  \subfloat[]{\label{fig:3ord_332B_2}\includegraphics[scale=0.50]{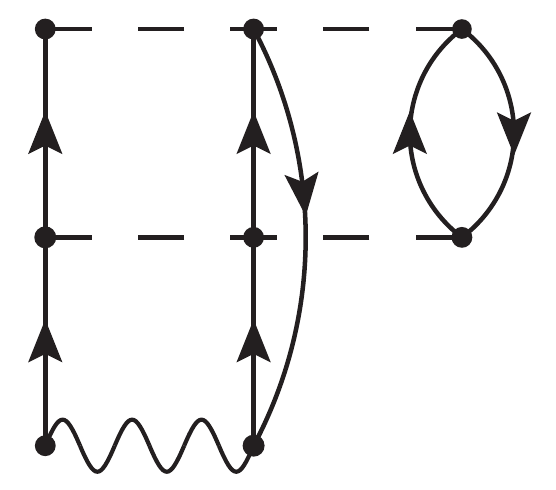}}
  \newline   \vskip .6cm
  \subfloat[]{\label{fig:3ord_323B_1}\includegraphics[scale=0.50]{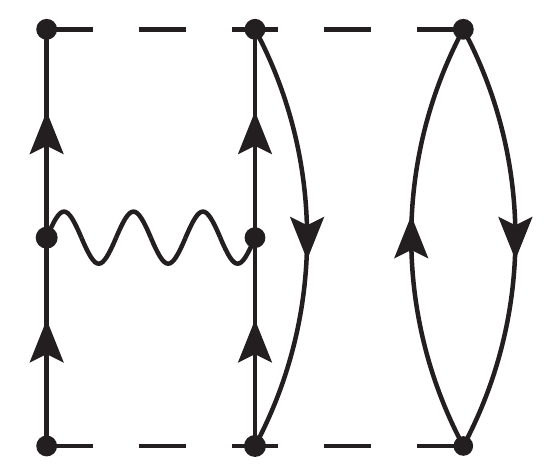}}
  \hspace{1.3cm}
  \subfloat[]{\label{fig:3ord_323B_2}\includegraphics[scale=0.50]{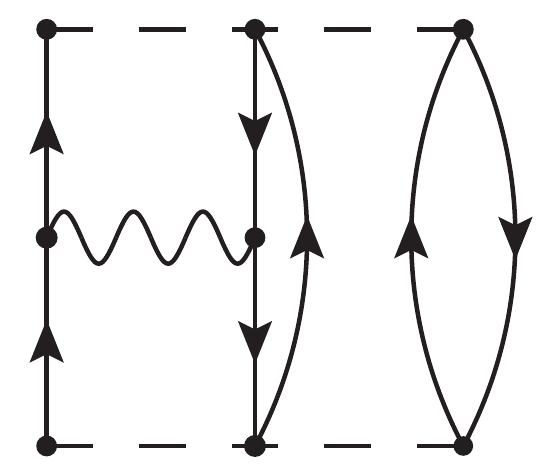}}
  \hspace{1.2cm}
  \subfloat[]{\label{fig:3ord_323B_3}\includegraphics[scale=0.50]{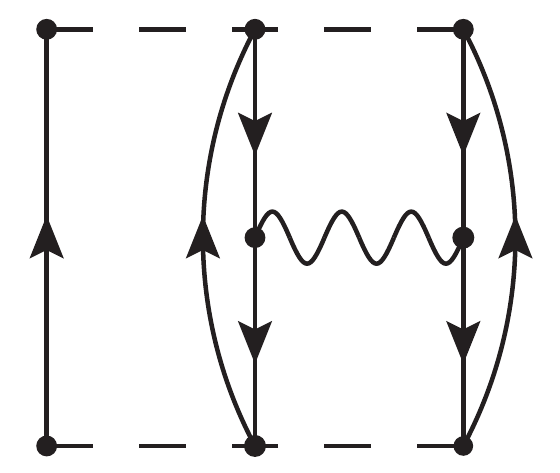}}
  \hspace{4cm}
  \newline   \vskip .6cm
  \hspace{3.cm}
  \subfloat[]{\label{fig:3ord_333B_1}\includegraphics[scale=0.55]{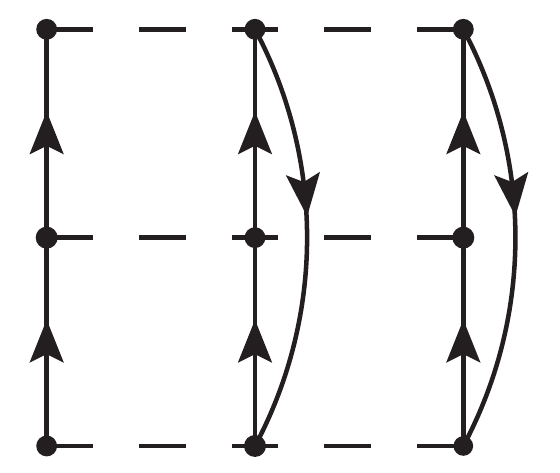}}
  \hspace{1.3cm}
  \subfloat[]{\label{fig:3ord_333B_2}\includegraphics[scale=0.55]{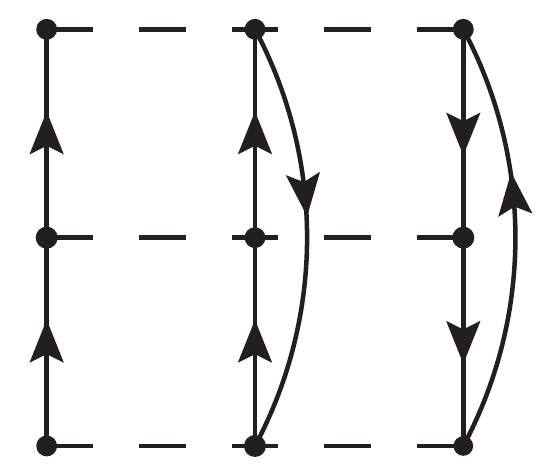}}
  \hspace{1.3cm}
  \subfloat[]{\label{fig:3ord_333B_3}\includegraphics[scale=0.55]{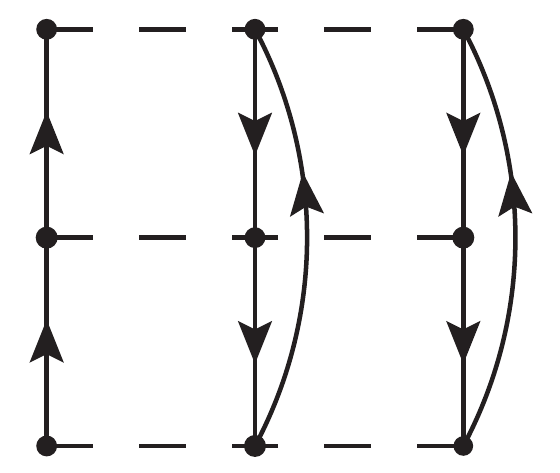}}
   \caption{ \small
   \emph{Interaction-irreducible} diagrams that contribute to the irreducible self-energy at third order.
   These are the only third-order contributions that must be considered when using the
   effective Hamiltonian of Eq.~(\ref{Heff}).} 
  \label{fig:3ord}
\end{figure}

At second order, the perturbative expansion of the irreducible self-energy generates five 
diagrams. By exploiting the two-body effective interaction $\widetilde{V}$, these are all grouped
into the two interaction-irreducible terms show in Fig.~\ref{fig:2ord}.
Likewise, there exist 53 diagrams at third order which reduce to the 17 interaction-irreducible
ones shown in Fig.~\ref{fig:3ord}.

Although they have not all been investigated in actual calculations, the diagrams of
Figs.~\ref{fig:2ord} and~\ref{fig:3ord} are expected to contribute very differently to
the self-energy and other computed quantities.
In these figures, they are ordered in terms of their expected decreasing importance.
Figs.~\ref{fig:2ord_2B},~\ref{fig:3ord_2B_1} and ~\ref{fig:3ord_2B_2} are the most relevant 
terms and are those already appearing in the case of only NN interactions. However,
they account here for the effects of 3NFs at lowest order through the effective NN
term of Eq.~\eqref{veff}. These all involve $2p1h$ and $2h1p$ intermediate states
and are the first contributions to the ADC(3) and FRPA expansions.
Diagram~\ref{fig:3ord_232B} does not introduce any new intermediate state but describes
the mixing among $2p1h$ and $2h1p$ configuration due to irreducible 3NFs. This
is expected to be less important on the basis that 3NFs are generally weaker than the
corresponding NN force (typically, $<\widehat{W}>\approx\f{1}{10}<\widehat{V}>$ for nuclear
interactions~\cite{Grange1989,Epelbaum2009rmp}).  Studies of normal-ordered Hamiltonians
also clearly suggest that this term gives a small correction to the total energy compared
to  diagrams~\ref{fig:3ord_2B_1} and~\ref{fig:3ord_2B_2}~\cite{Hagen2007cc3nf,Roth2012prl}.
Note that diagrams~\ref{fig:3ord_2B_1}-\ref{fig:3ord_232B} contain the first-order terms
of the all order summation that generates configuration mixing between $2p1h$ or
$2h1p$ excitations.  Nowadays, resummations of these configurations are performed routinely
for the \ref{fig:3ord_2B_1} the \ref{fig:3ord_2B_2} terms in ADC(3)
and FRPA calculations~\cite{Barbieri:2007Atoms,Barbieri2009ni,Ortiz2013}.

The diagrams in Figs.~\ref{fig:2ord_3B} and~\ref{fig:3ord_223B_1} to \ref{fig:3ord_332B_2} introduce
additional configurations of the  $3p2h$ and $3h2p$ type.  These are expected to 
affect sensibly the distribution of small fragments in the particle and hole spectral
strength distribution but have smaller relevance for calculation of
total binding energies and dominant quasipartcle and quasihole peaks.
All these third order contributions provide the transition matrix elements connecting
$2p1h$ to $3p2h$ (and $2h1p$ to $3h2p$). However,
diagrams~\ref{fig:3ord_223B_1} to~\ref{fig:3ord_322B_2} may be more important than the remaining
four, since they involve one more NN vertex.
Note that the terms~\ref{fig:2ord_3B} and~\ref{fig:3ord_223B_1} to~\ref{fig:3ord_332B_2}
together induce the same type of configurations that appear only from the fourth-order ADC(4) and beyond, when 3NFs are missing. Thus, they are expected to have similar impact on the many-body truncation.
The remaining diagrams, \ref{fig:3ord_323B_1} to~\ref{fig:3ord_333B_3}, describe the explicit
mixing among $3p2h$ or $3h2p$ configurations and are expected to give contributions 
comparable to the NN-only ADC(5) approximation.

The energy sum rule originally proposed by Galitskii-Migdal~\cite{Galitskii1958KSR} and by Koltun~\cite{Koltun1972KSR} needs to be extended for the presence of many-body interactions. When only up to three-body forces are present, this can be re-expressed in one of two equivalent forms:
\begin{equation}
\label{eq:koltunW}
E^A_0 = \sum_{\alpha \, \beta} \frac{1}{4 \pi i} \int_{C \uparrow} d \omega  \; \;
\left[ U_{\alpha \beta}  +  \omega \delta_{\alpha \beta} \right] \, g_{\beta \alpha}(\omega)   
~ - ~ \frac{1}{2}\langle {\Psi^A_0}     \vert W     \vert {\Psi^A_0} \rangle  
\end{equation}
and
\begin{equation}
\label{eq:koltunV}
E^A_0 = \sum_{\alpha \, \beta} \frac{1}{6 \pi i} \int_{C \uparrow} d \omega  \; \;
\left[2 U_{\alpha \beta}  +  \omega \delta_{\alpha \beta} \right] \, g_{\beta \alpha}(\omega)   
~ + ~ \frac{1}{3}\langle {\Psi^A_0}     \vert V     \vert {\Psi^A_0} \rangle  \; .
\end{equation}
Both Eqs.~\eqref{eq:koltunW} and~\eqref{eq:koltunV} are exact but they rely on the possibility of estimating the expectation value of either the two- or the three-body part of the Hamiltonian. Most recent SCGF calculations exploited Eq.~\eqref{eq:koltunW} based on the consideration that absolute errors in evaluating $\bra{\Psi^A_0} \hat{W} \ket{\Psi^A_0} $ should be smaller, since the contribution of the 3NF is about an order of magnitude less than that of the NN force~\cite{Cipollone2013prl,Carbone2013snm,Soma2014s2n}.
 For finite nuclei and SRG cutoffs of $\lambda_{SRG}$=2.0~fm$^{-1}$, we found that  $\bra{\Psi^A_0} \hat{W} \ket{\Psi^A_0}$ can be evaluated at first order in the {\em dressed} propagators and this is sufficient to converge the calculated binding energy with respect to many-body truncations~\cite{Cipollone2013prl}. This is the same level of approximation needed to converge the effective potential $\widetilde{U}$ with respect to many-body truncations (see Fig.~\ref{fig:Ueff_conv} and its discussion above).

\section{Gorkov formulation for semi-magic open shell systems} ~

In Refs.~\cite{Soma:2011GkvI, Soma:2013rc, Soma2014GkvII}, we have introduced an extension of the SCGF scheme that is based on the Gorkov's formalism \cite{Gorkov:1958} and that can address open shell nuclei.
 The Gorkov method is based on the idea of allowing the breaking of particle number symmetry in order to achieve an effective description of the pairing correlations, which play a crucial role in open shell systems and remove degeneracies in the reference state.
 In order to work with the correct number of protons and neutrons in average, chemical potentials are  added to the Hamiltonian,
that is, we consider the grand canonical  Hamiltonian $\hat\Omega(A) = H - \hat{T}_{\text{c.o.m.}}(A) -  \mu_{\text{p}} \, \hat{Z} - \mu_{\text{n}} \, \hat{N}$.
Additional (anomalous) propagators that account for the breaking and formation of Cooper pairs are introduced.
The Lehman representation of the Gorkov propagators is
\begin{subequations}
\label{eq:Gab}
\begin{eqnarray}
\label{eq:Gab11}
G^{11}_{\alpha \beta} (\omega) &=&  \sum_{k} \left\{
\frac{\mU_{\alpha}^{k} \,\mU_{\beta}^{k*}}
{\omega-\omega_{k} + \ii \eta}
+ \frac{\bar{\mV}_{\alpha}^{k*} \, {\bar{\mV}_{\beta}^{k}}}{\omega+\omega_{k} - \ii \eta} \right\} \: ,\\
\label{eq:Gab12}
G^{12}_{\alpha \beta} (\omega) &=&   \sum_{k}
\left\{
\frac{\mU_{\alpha}^{k} \,\mV_{\beta}^{k*}}
{\omega-\omega_{k} + \ii \eta} + \frac{\bar{\mV}_{\alpha}^{k*} \, {\bar{\mU}_{\beta}^{k}}}{\omega+\omega_{k} - \ii \eta}
\right\}  \, ,\\
\label{eq:Gab21}
G^{21}_{\alpha \beta} (\omega) &=&  \sum_{k}
\left\{
\frac{\mV_{\alpha}^{k} \,\mU_{\beta}^{k*}}
{\omega-\omega_{k}  + \ii \eta}
+ \frac{\bar{\mU}_{\alpha}^{k*} \, {\bar{\mV}_{\beta}^{k}}}{\omega+\omega_{k} - \ii \eta}
\right\} \, ,\\
\label{eq:Gab22}
G^{22}_{\alpha \beta} (\omega) &=&   \sum_{k} \left\{
\frac{\mV_{\alpha}^{k} \,\mV_{\beta}^{k*}}
{\omega-\omega_{k} + \ii \eta}
+  \frac{\bar{\mU}_{\alpha}^{k*} \, {\bar{\mU}_{\beta}^{k}}}{\omega+\omega_{k} - \ii \eta} \right\} \: ,
\end{eqnarray}
\end{subequations}
where the poles of the propagators are now given by \hbox{$\omega_{k} \equiv \Omega_k - \Omega_0$},  $\Omega_0$ is the ground state of the even-A isotope and the index $k$ labels odd-A eigenvalues 
\hbox{of $\hat{\Omega}$:~$\hat{\Omega} \, | \Psi_{k} \rangle = \Omega_{k} \, | \Psi_{k} \rangle$.} 
The residues of pole $\omega_{k}$ relate to the probability amplitudes 
$\mU^k$ ($\mV^k$) to reach state $| \Psi_{k} \rangle$ by adding (removing) a nucleon to (from) the even-A ground state $| \Psi_{0} \rangle$.

 Formally, Eqs. \eqref{eq:Dyson} and \eqref{eq:selfenergy} still hold with all quantities (propagators and self-energies) now being  matrices in a $2 \times 2$ Gorkov space. 
 The self-energy still splits into static and dynamic contributions, $\Sigma^{g_1 g_2}(\omega)=\Sigma^{g_1 g_2, (\infty)}+\Sigma^{g_1 g_2 , (dyn)}(\omega)$~(where $g_i$=1,2 are the Gorkov indices).
The working equations for the Gorkov formalism have been derived in full up to second order and NN interactions~\cite{Soma:2011GkvI}. The corresponding self-energy diagrams are shown Figs.~\ref{fig:se_exp}b to~\ref{fig:se_exp}e.
In the present implementation of the Gorkov formalism, we add 3NFs as described in Sec.~\ref{sec:3nf} by exploiting the effective interactions of Eqs.~\eqref{ueff} and~\eqref{veff} and adding the anomalous term corresponding to $\tilde{U}$~\cite{Soma2014s2n}.
The extension of the Gorkov framework to FRPA and ADC(3) type of self-energies is currently underway.

\section{Studies of correlations and optical potentials} \label{SecOptMod} ~

In Eq.~\eqref{eq:Dyson}, the irreducible self-energy $\Sigma^\star(\omega)$---or, equivalently, the Gorkov's normal self-energy $\Sigma^{11}(\omega)$---acts as an energy dependent non-local single particle potential. In fact, it can be proven that $\Sigma^\star(\omega)$ is exactly the Fesbach microscopic optical potential, once this is extended to treat both nucleon-nucleus states (above the Fermi surface, $E_F$) and hole-nucleus states (below $E_F$)~\cite{Capuzzi96,Escher2002opt}. This provides a way to calculate optical potentials from {\em ab-initio} theory.

In order to investigate the microscopic properties of optical models, we transform Eq.~\eqref{eq:selfenergy} to coordinate space and separate its central and spin-orbit angular momentum components as follows~\cite{Waldecker:2011by}:
\begin{eqnarray}
\Sigma^\star( \bm{x}, \bm{x^\prime}; E ) 
&=& \sum_{\ell j m_j \tau} {\cal I }_{\ell j m_j}( \Omega, \sigma ) 
 \left[ \sum_{n_a, n_b} R_{n_a \ell}(r) \Sigma^\star_{ab}(E)R_{n_b \ell}( r^\prime )\right] ( {\cal I }_{\ell j m_j}( \Omega^\prime, \sigma^\prime ) )^* 
\nonumber \\
&=& \sum_{\ell j m_j \tau} {\cal I }_{\ell j m_j}( \Omega, \sigma ) 
 \left[
 \Sigma^{\ell}_{0}(r, r'; E)+ (\vec{\bm{\ell}} \cdot \vec{\bm{\sigma}}) \, \Sigma^{\ell}_{\ell s}(r, r'; E)
 \right] ( {\cal I }_{\ell j m_j}( \Omega^\prime, \sigma^\prime ) )^* ,
\label{eq:selfr}
\end{eqnarray}
where $\bm{x} \equiv \bm{r}, \sigma, \tau$.  
In \eqref{eq:selfr}, $\sigma$ represents the spin variable, $\tau$ is the isospin, $n$ is the principal quantum number of the  harmonic oscillator basis used in the calculations, and $a\equiv(n_a, \ell , j, \tau)$ (note that for a $J = 0$ target the self-energy is independent of $m_j$).
The standard radial harmonic-oscillator function is denoted by $R_{n \ell}(r)$, while ${\cal I }_{\ell j m_j}( \Omega, \sigma )$ represents the $j$-coupled angular-spin function. We calculated the potential \eqref{eq:selfr} in Ref.~\cite{Waldecker:2011by} by using FRPA and the Argonne V18 NN interaction~\cite{Wiringa95}.

The most recent implementations of the dispersive optical model (DOM) are substantially phenomenological optical potentials whose analytical structure is, however, constrained by our best knowledge of ab-initio calculated self-energies~\cite{Charity2006prl,Charity2007dom,Mueller2011dom,Mahzoon2014prl}. In fact, the DOM can be seen as a phenomenological parameterisation of $\Sigma^\star(\omega)$ that allows for data driven extrapolations to exotic isotopes. 
 Note that the analytic structure of Eq.~\eqref{eq:selfenergy} implies a dispersive relation that links the real and imaginary parts of $\Sigma^\star(\omega)$ and that it encodes the causality principle. This relation is at the core of the DOM approach. 
In fitting optical potentials, it is usually found that volume integrals are better constrained by the experimental data~\cite{Mahaux91,Greenless68}. Thus, we focus our attention on the imaginary part of the central self-energy, which corresponds
to the total absorption. Its volume integral is given by
\begin{equation}
J_W^\ell(E) = 4\pi\int{drr^2\int{dr^\prime r^{\prime 2} \; \text{Im } \Sigma^{\ell}_0(r, r^\prime ; E)}} \, .
\label{eq:intgs_W}
\end{equation}

\begin{figure}[t!]
\begin{center}
\includegraphics[height=0.57\textwidth,width=0.49\textwidth,clip]{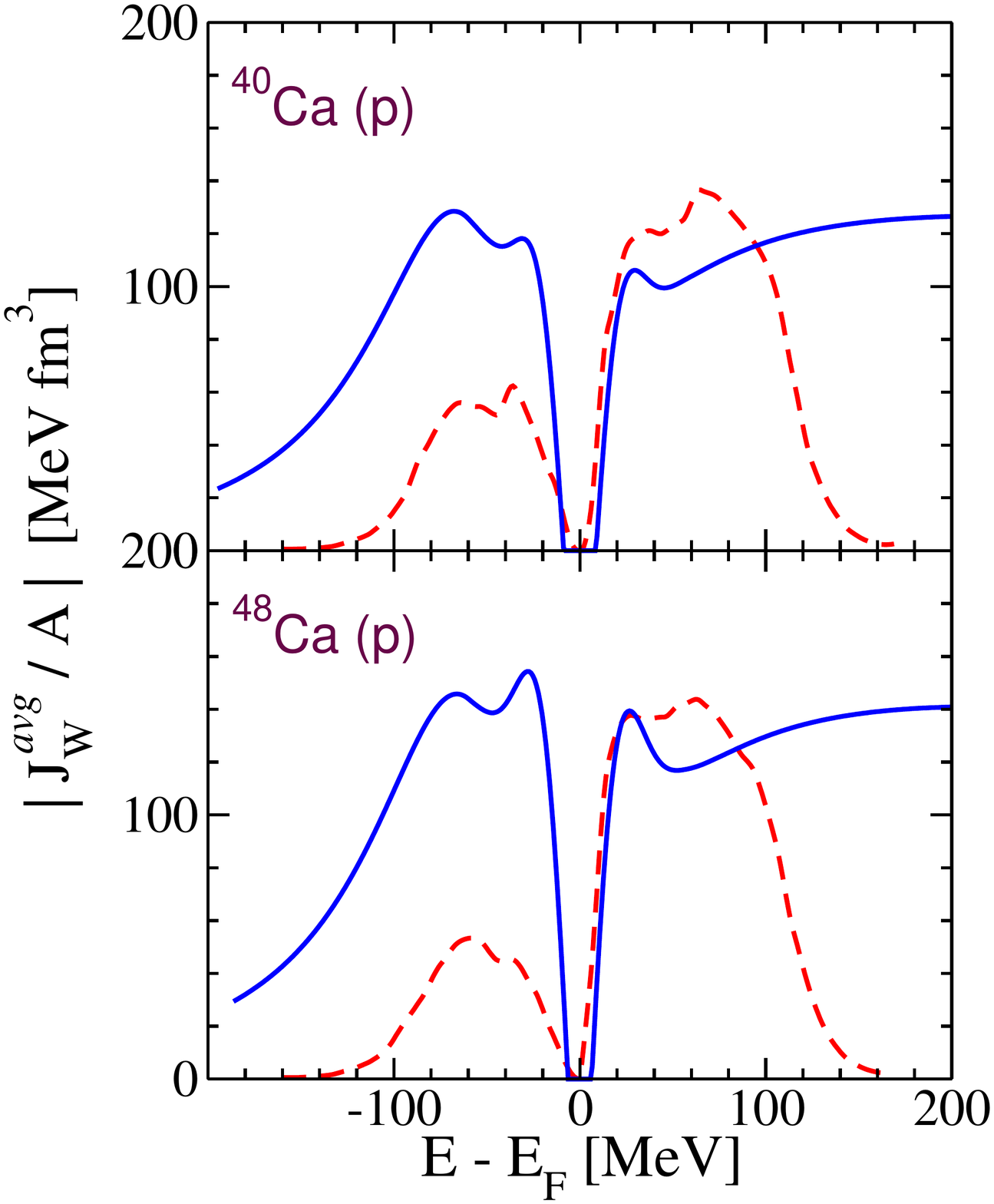}
\hspace{.2 cm}
\includegraphics[width=0.47\textwidth,clip]{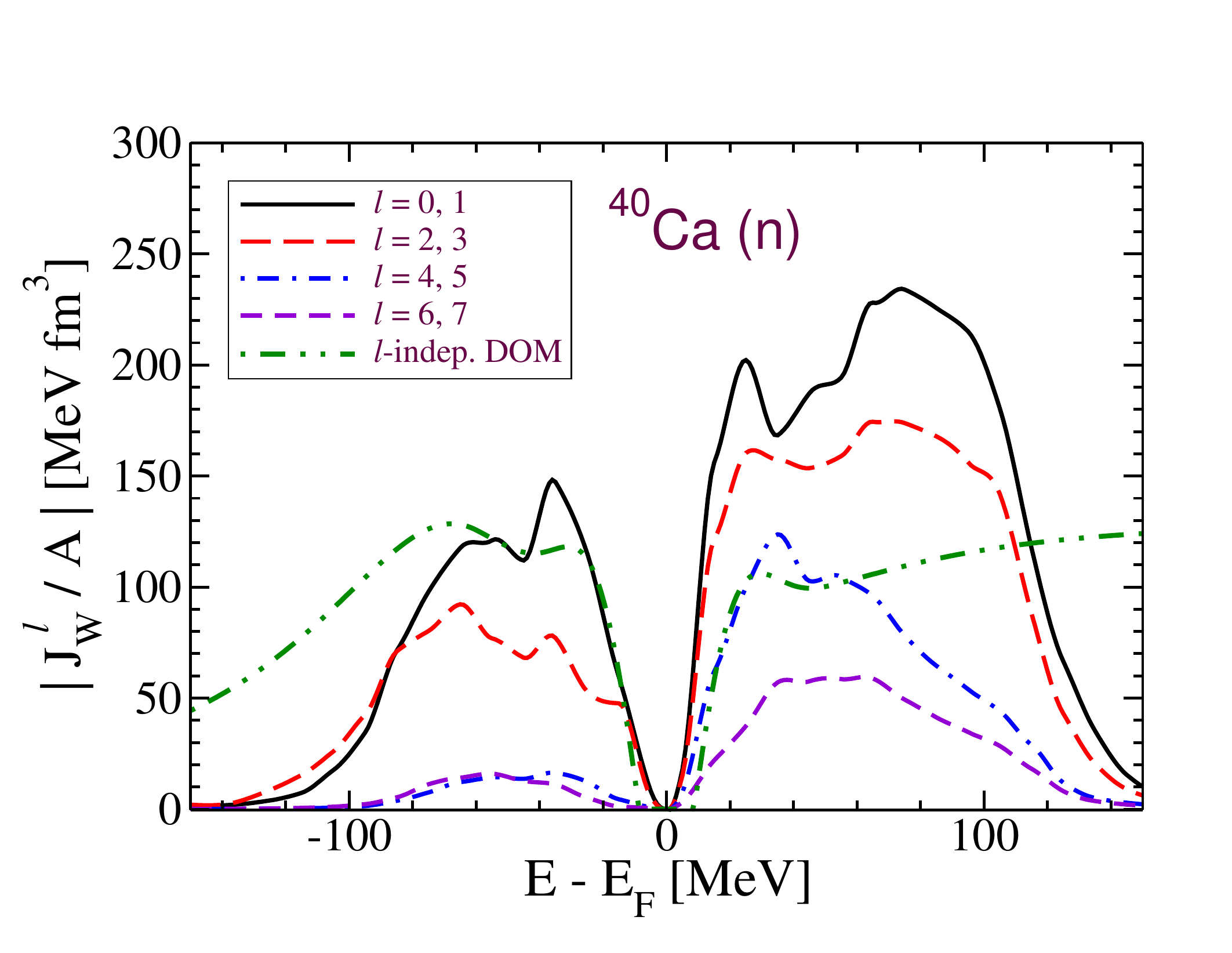}
\vskip -.2cm
\caption{  \small
{\em Left}:  The FRPA results for the volume integral averaged over all $\ell$-channels $J_w^{avg}$(dashed line) are compared with the DOM result (solid line), corrected for non-locality, for proton scattering on~$^{40}$Ca and~$^{48}$Ca.
{\em Right}: Separate partial wave contributions to $J_W$ averaged over adjacent $\ell$-channels in the model space. This plot is for neutrons in $^{40}$Ca. The dash-double-dotted curve represents the DOM result.
All calculation are based on the NN-only AV18 realistic interaction.
}
\vspace{-0.5cm}
\label{fig:Pdep}       
\end{center}
\end{figure}

The left panel of Fig.~\ref{fig:Pdep} compares the $\ell$-averaged FRPA volume integrals ($J_w^{avg}$) with the corresponding global DOM potentials of Ref.~\cite{Charity2007dom}. This local form of the DOM has been corrected by the effective mass~\cite{Mahaux91,Dickhoff2011nonloc} to account for non-locality. The overall effect of this correction is to enhance the absorption. 
The FRPA exhibits different behavior above and below $E_F$ than the DOM. Microscopic calculations predict significantly less absorption below $E_F$, whereas in the DOM fit, the absorption was assumed to be roughly symmetric up to about 50 MeV away from $E_F$~\cite{Mahaux91,Charity2007dom,Mueller2011dom}.
%
Since only the absorption above the Fermi energy is strongly constrained by elastic scattering data, it is encouraging that the $\ell$-averaged FRPA result is reasonably close to the DOM fit for both nuclei, up to $E-E_F <$ 80~MeV where the FRPA results are not sensibly affected by the truncation of the model space.

The simplifying assumptions of a symmetric absorption around $E_F$ and locality in the DOM imply an unrealistic occupation of higher $\ell$-values below the Fermi energy which is not obtained in the FRPA. This is shown  in the right panel of Fig.~\ref{fig:Pdep} for scattering of neutrons against ${}^{40}$Ca. 
Below the Fermi energy the contribution from {\em s}, {\em p} and {\em sd} orbits dominates while higher $\ell$-values are suppressed. This is a clear indication of the nucleon occupations in the mean field shell structure of the target nucleus. The dash-double-dotted curve illustrates the same DOM fit also shown in the left panel. This compares favourably to FRPA for the low-$\ell$ partial waves below the Fermi surface.
Recent fits of the DOM include non-locality effects explicitly and find that this is required in order to reproduce the features of spectral distribution of single particle strength below $E_F$, including the experimentally observed position of quasiparticle peaks~\cite{Mahzoon2014prl}. This also allow for the description of high-momentum components that is suggested by ($e$,$e'p$) observations at Jefferson Laboratory~\cite{Rohe2004prl,Barbieri2004rescatt,Barbieri2005plb}.

\begin{figure}[t]
\begin{center}
\includegraphics[width=0.56\textwidth,clip]{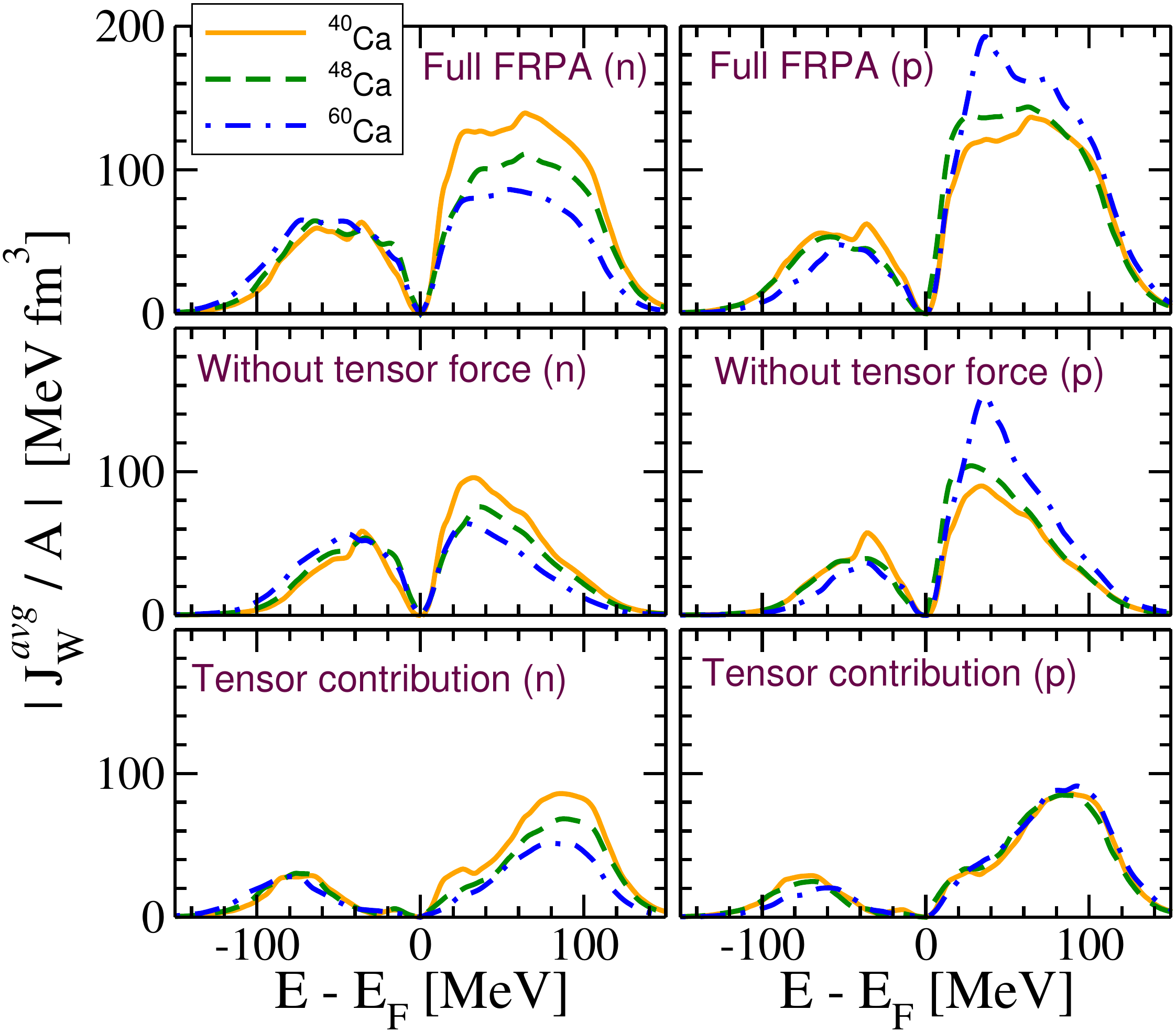}
\hspace{.3 cm}
\includegraphics[width=0.39\textwidth,clip]{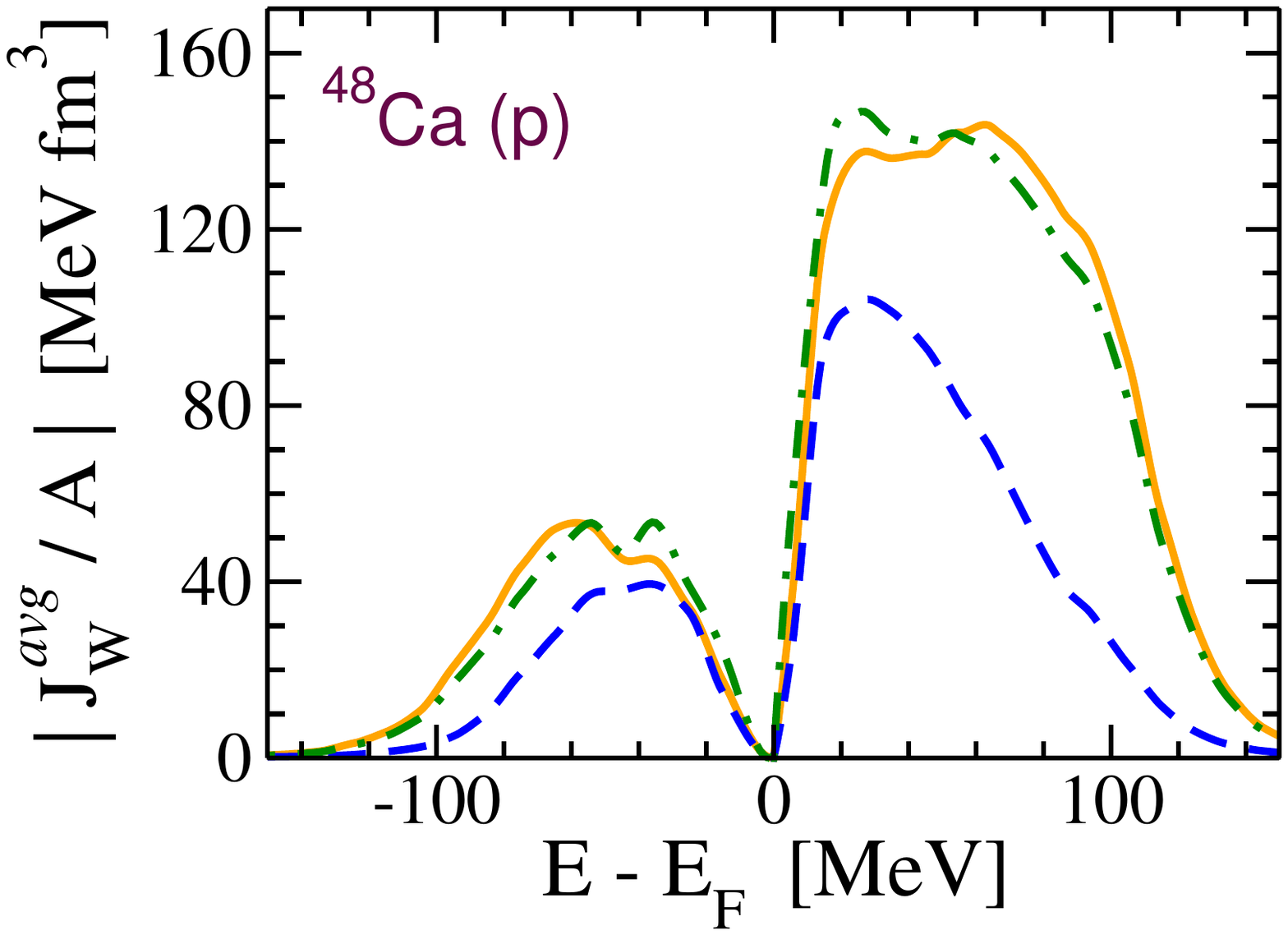}
\caption{ \small
{\em Left}:  Asymmetry dependence of the absorption for neutrons and protons and dependence on tensor correlations.  $J_W^{avg}$ for $^{40}$Ca (solid) is compared with the results for $^{48}$Ca (dashed), and $^{60}$Ca (dot-dashed).
The figures in the lowest row are the difference between the full calculation and the middle row figures, to show the effect of correlations arising from the tensor force.
{\em Right}: Effect of the tensor force and charge exchange correlations on the proton-$^{48}$Ca self-energy. The solid curve is the imaginary volume integral $J_W^{avg}$ from the full FRPA calculation, while the dashed curve results from removing the tensor term in the AV18 interaction. The dash-dotted curve is obtained by excluding charge exchange from the full calculation.  Similar results are found for neutrons and the other Ca isotopes.
All calculation are based on the NN-only AV18 realistic interaction.
}
\label{fig:asymm_tensor_dep}       
\end{center}
\end{figure}

 The behavior of the nuclear self-energy with changing proton-neutron asymmetry \hbox{($\alpha = (N-Z)/A$)} is very important in seeking for  {\em global} parametrizations of the DOM. Hence its understanding is pivotal in predicting unstable isotopes. 
 The changes in the calculated $J_W^{avg}$ for isotopes going from $^{40}$Ca to $^{60}$Ca are shown in the top-left panels of Fig.~\ref{fig:asymm_tensor_dep}. These cover a range of asymmetries form $\alpha$=0 to $1/3$.
 Microscopic FRPA predicts an opposite behavior of protons and neutrons above $E_F$, with the proton (neutron) potential increasing (decreasing) when more neutrons are added, qualitatively in agreement with expectations from the Lane potential model~\cite{Lane62}. The DOM fit of Ref.~\cite{Mueller2011dom} includes several isotopes in the Ca and Sn chains and employs a similar trend for the volume integrals.

The characteristics of the above asymmetry dependence indicate that the nuclear interaction between protons and neutrons plays a major role.   The tensor force of the nuclear interaction can provide one such mechanism since it is particularly strong in the isospin $T=0$ sector. Moreover, it has already been shown to influence the evolution of single-particle energies at the Fermi surface~\cite{Otsuka05prl}.
To investigate its implication for single-particle properties at energies farther removed from $E_F$, we recalculated $J_W^{avg}$ by suppressing the tensor component of the AV18 interaction. 
It is apparent that this has a very significant effect on the correlations far from the Fermi surface for nucleon-nucleus scattering. On the other hand, absorption near the Fermi surface are dominated by correlations other than tensor.

In the right panel of Fig.~\ref{fig:asymm_tensor_dep} (dot-dashed line), we have also calculated the FRPA self-energy by suppressing charge-exchange excitations in the polarization propagator $\Pi^{(ph)}$ (see Fig.~\ref{fig:se_exp}a). These contributions correspond to a mechanism in which the proton (neutron) projectile is exchanged with a neutron (proton) in the target. This includes Gamow-Teller resonances whose strength increase with asymmetry as $\approx 3(N-Z)$~\cite{Charity2007dom}. Ab-initio FRPA calculations indicate that such correlations do not generate significant absorption in scattering processes.

\section{Role of chiral three-nucleon forces in medium mass isotopes}  \label{Sec3NF} ~

Calculations were performed using chiral NN and 3N forces evolved to low momentum scales through free-space similarity
renormalization group (SRG) techniques~\cite{Jurgenson2009prl}.
The original NN interaction is generated at next-to-next-to-next-to-leading order (N$^3$LO) with cutoff $\Lambda_{NN}$=500 MeV~\cite{Entem2003N3LO,Machleidt2011pr} and is supplemented by a local NNLO 3NF~\cite{Navratil2007tnf} with a reduced cutoff of $\Lambda_{3N}$=400~MeV, as in Ref.~\cite{Roth2012prl}. 
The chiral NNLO 3NF contain the two-pion exchange Fujita-Miyazawa contribution. 
The SRG evolution of the sole chiral NN interaction already generates 3N operators in the Hamiltonian, which we refer to hereafter as the ``induced'' 3NF. When the pre-existing chiral 3N interaction is also included, we refer to it as the ``full'' 3NF.
%
A SRG cutoff $\lambda_{SRG}$=2.0~fm$^{-1}$ was used for most calculations presented in this section.


\begin{figure}[t]
\begin{center}
\includegraphics[width=0.49\textwidth,clip]{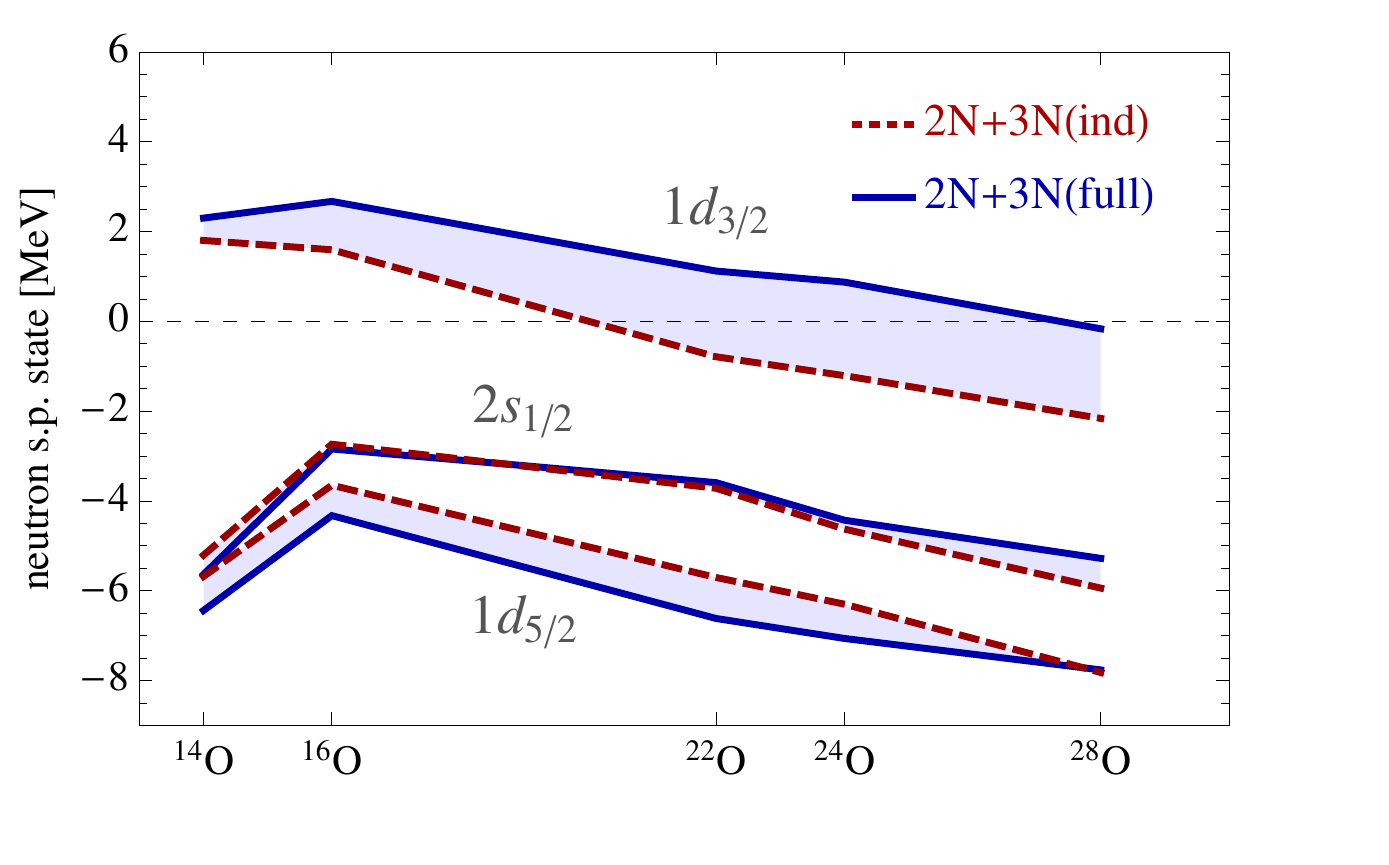}
\includegraphics[width=0.49\textwidth,clip]{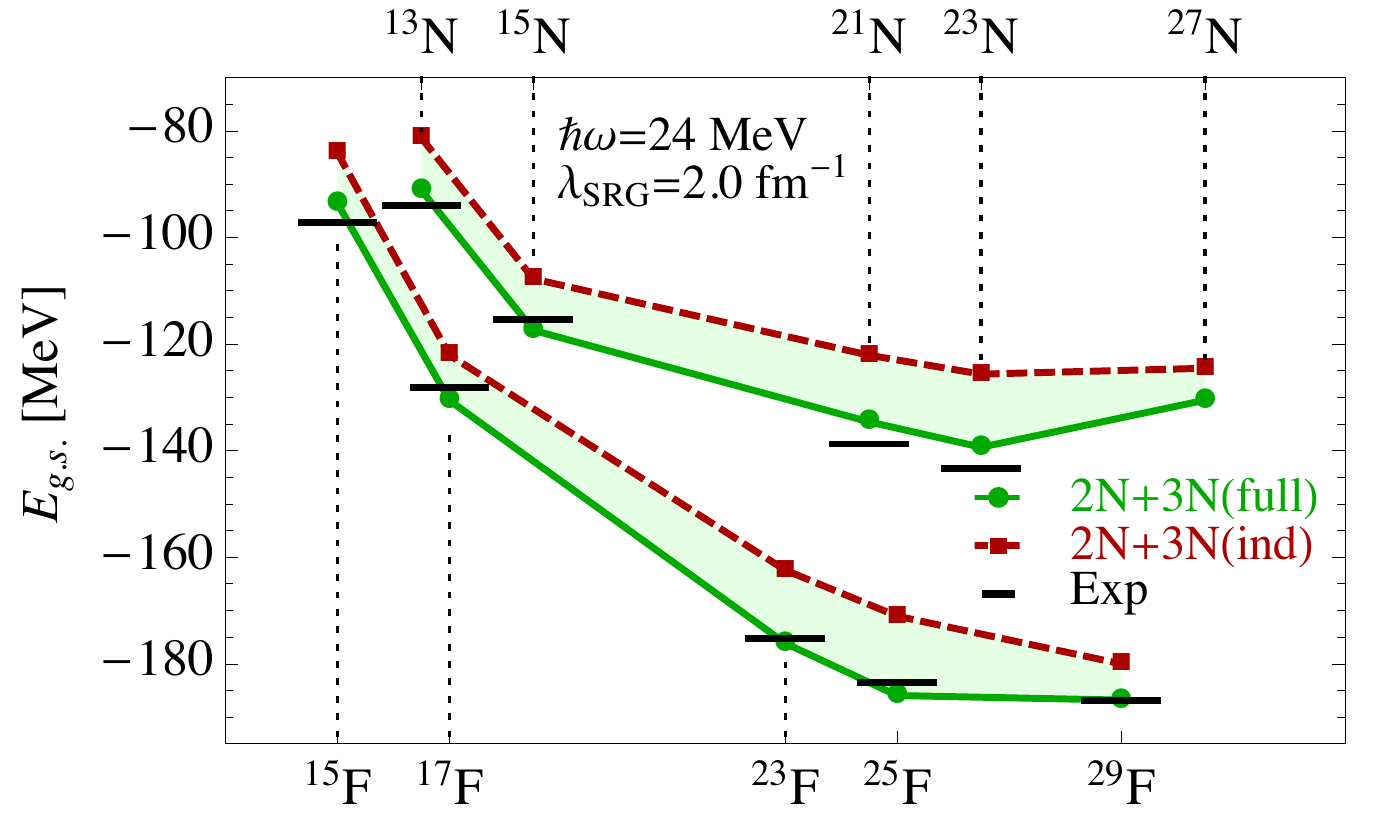}
\caption{ \small
{\em Left}: Energies for the addition and separation of a neutron to and from various oxygen isotopes. These correspond to the dominant quasiparticle peaks that can be observed in the final $Z$=8 and $N\pm$1 isotopes.
{\em Right}: Ground-state energies of nitrogen and fluorine isotopes calculated for induced and full interactions. Experimental data are from~\cite{AME2003,Jurado2007,Gaudefroy2012}.
Calculations have been performed with chiral two- and three-body interactions evolved to $\lambda_{SRG}=2.0 \, \text{fm}^{-1}$ by means of SRG techniques~\cite{Cipollone2013prl}. 
In both plots, the dashed and full lines join results for the induced and full interactions, respectively.
}
\label{fig:NFenergies}       
\end{center}
\end{figure}

Fully ab-initio calculations based on evolved NN plus full 3NFs can accurately reproduce ground state energies of the entire oxygen isotopic chain, as shown in Fig.~\ref{fig:Ueff_conv} and reported in Ref.~\cite{Barbieri:2012rd}. This was subsequently confirmed by extended calculations across different many-body methods, which estimated in details the errors associated with the SRG evolution and the many-body truncations schemes~\cite{Hergert2013prl,Cipollone2013prl}.  These results give a first principle confirmation of the repulsive effects of the two-pion exchange Fujita-Miyazawa interaction, which was found to be at the origin of the anomalous dripline at $^{\rm 24}$O~\cite{Otsuka2010prl}.
This is seen in the left panel of Fig.~\ref{fig:NFenergies}, which shows the predicted evolution of neutron single particle spectrum (addition and separation energies) for oxygen isotopes. Induced 3NFs reproduce the overall trend but predict a bound $d_{3/2}$ when the  {\em sd} shell is filled.  Adding pre-existing 3NFs---the full Hamiltonian---raises this orbit above the continuum also for the highest masses.

The same mechanism affects neighboring isotopic chains, as demonstrated in the right panel of Fig.~\ref{fig:NFenergies} for the odd-even semi-magic nitrogen and fluorine. Induced 3NF forces consistently  under bind these isotopes and even predict a $^{27}$N close in energy to $^{23}$N. 
This is fully corrected by full 3NFs that strongly binds $^{23}$N with respect to $^{27}$N, in accordance with the experimentally observed dripline.  The repulsive effects of filling the $d_{3/2}$ is also observed in $^{29}F$ but it is counter balanced by attractions due to the inclusion of an extra proton.  This leads to a slightly bound $^{29}F$ in accordance with observation~\cite{Gaudefroy2012}. 
Eventually, leading-order (NNLO) original three-body terms are crucial both to bring calculated energies close to the experiment and to yield a correct description of the drip lines.

%
%

\begin{figure}[t!]
\begin{center}
\includegraphics[width=0.4\textwidth,clip]{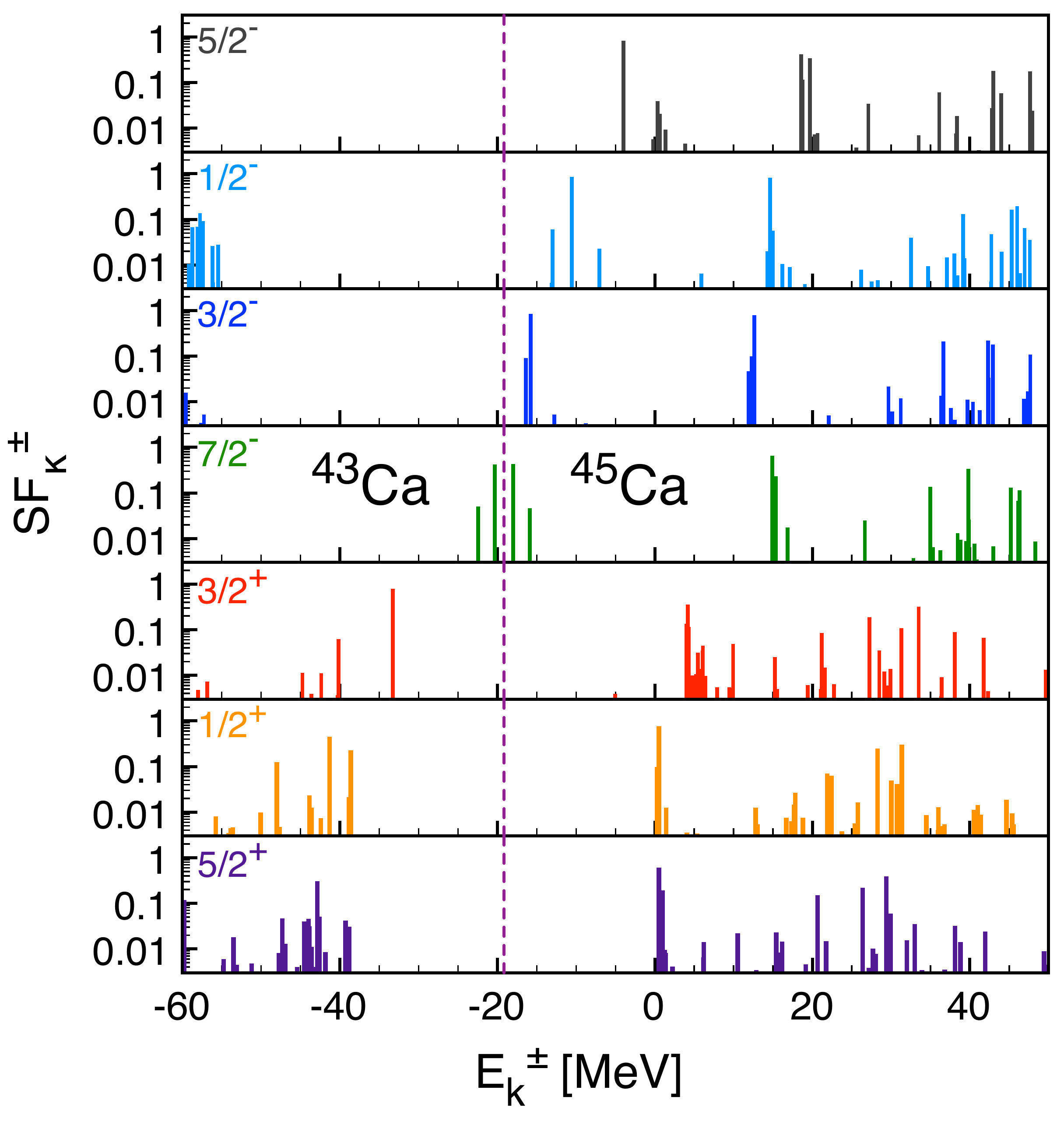}
\hspace{.7 cm}
\includegraphics[width=0.45\textwidth,clip]{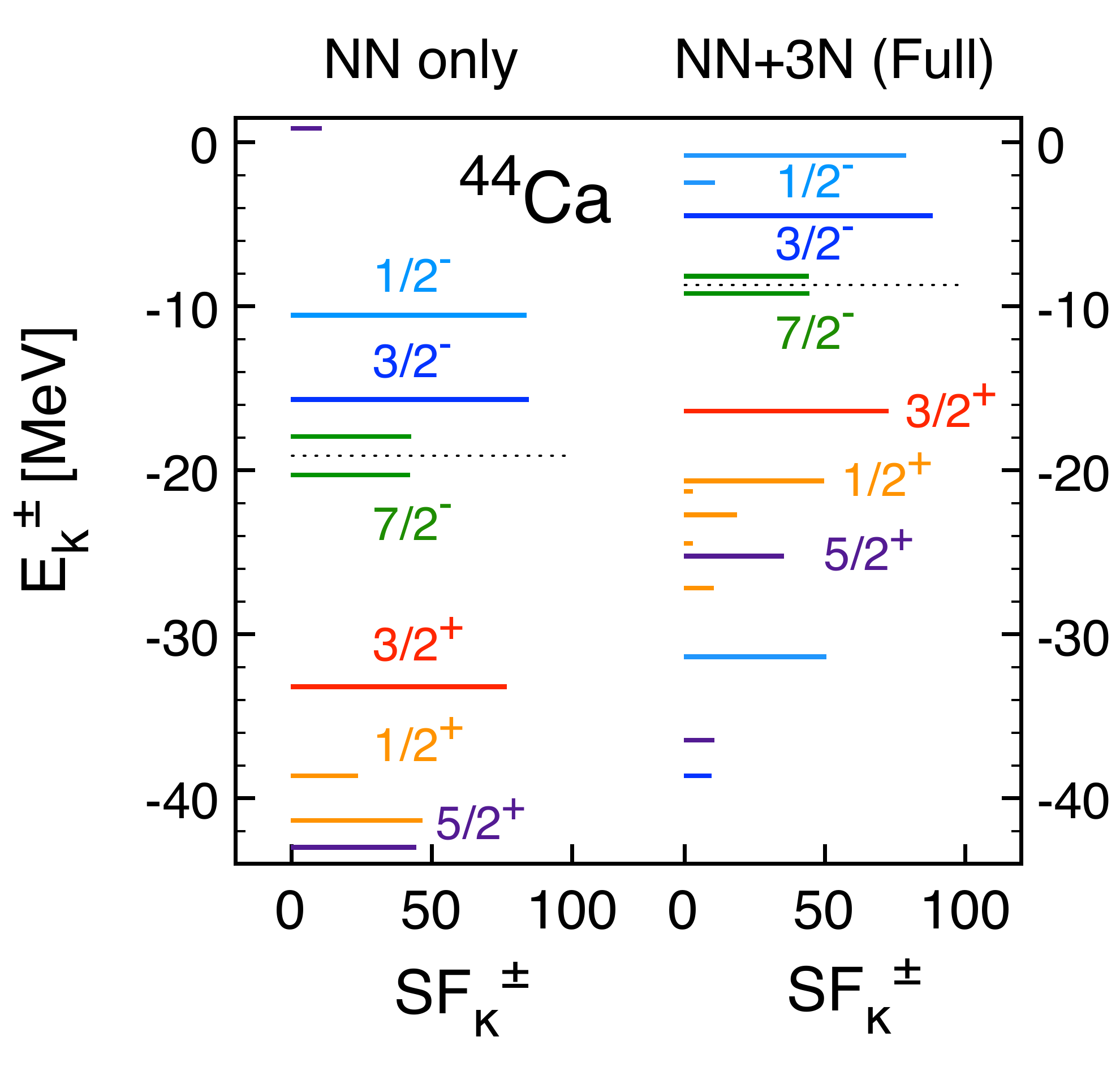}
\caption{\small Spectral strength distribution for the removal and addition of a neutron from and to $^{\rm 44}$Ca, calculated using Gorkov-GF at second order~\cite{Soma:2013rc}.
{\em Left}: Distribution of different partial waves calculated with an evolved NN interaction only. 
{\em Right}: Distribution in the energy region of the {\em sd} and {\em pf} nuclear shells, obtained from NN-only and 
NN plus full 3NFs.  The experimental neutron chemical potential (corresponding to the Fremi surface) is $\mu_n(^{\rm 44}{\rm Ca})$=-9.27~MeV and it is well reproduced by full 3NFs.
}
\label{fig:Ca44_SF}       
\end{center}

\end{figure}

Both Dyson and Gorkov formalisms yield the single particle spectral function. Its diagonal part,
\begin{equation} 
 S_{\al}(\omega) = \frac 1\pi | {\rm Im} \; g_{\al \al}(\omega) |  = \frac 1\pi | {\rm Im} \;  G^{11}_{\al \al}(\omega + \mu_\al) |  \; ,
\end{equation} 
describes the energy distribution of  the  spectral strength (i.e. spectroscopic factors) for the addition or removal of one particle.
This is demonstrated in Fig.~\ref{fig:Ca44_SF} for second-order Gorkov calculations of $^{\rm 44}$Ca. The left panel shows the spectral distribution for different partial waves. States below (above) the Fermi surface refer to the separation (addition) of a neutron, leading to final states of $^{\rm 43}$Ca~($^{\rm 45}$Ca). These results refer to calculations based on evolved NN interactions only, with all (induced and full) 3NFs discarded. The effects of 3NFs on the spectral distribution in the {\em sd} and {\em pf} shells is shown in the right panel, and it is compared to the NN-only results from the left. 
We find that the NN plus full 3NFs Hamiltonian is fundamental to predict the correct  location of the Fermi surface with respect to the single-particle continuum. 3NFs also reduce the average energy separation of the two major valence shell, {\em sd} and {\em pf}, by a factor of two and bring these in closer agreement with the experiment. In our calculations we find that the neutron separation energy to the $J^\pi=3/2^+$ state of $^{\rm 43}$Ca is about -16~MeV. This is close to the phenomenological predictions based on the DOM of Ref.~\cite{Charity2007dom}.
%
\begin{figure}[t!]
\begin{center}
\includegraphics[width=0.48\textwidth,clip]{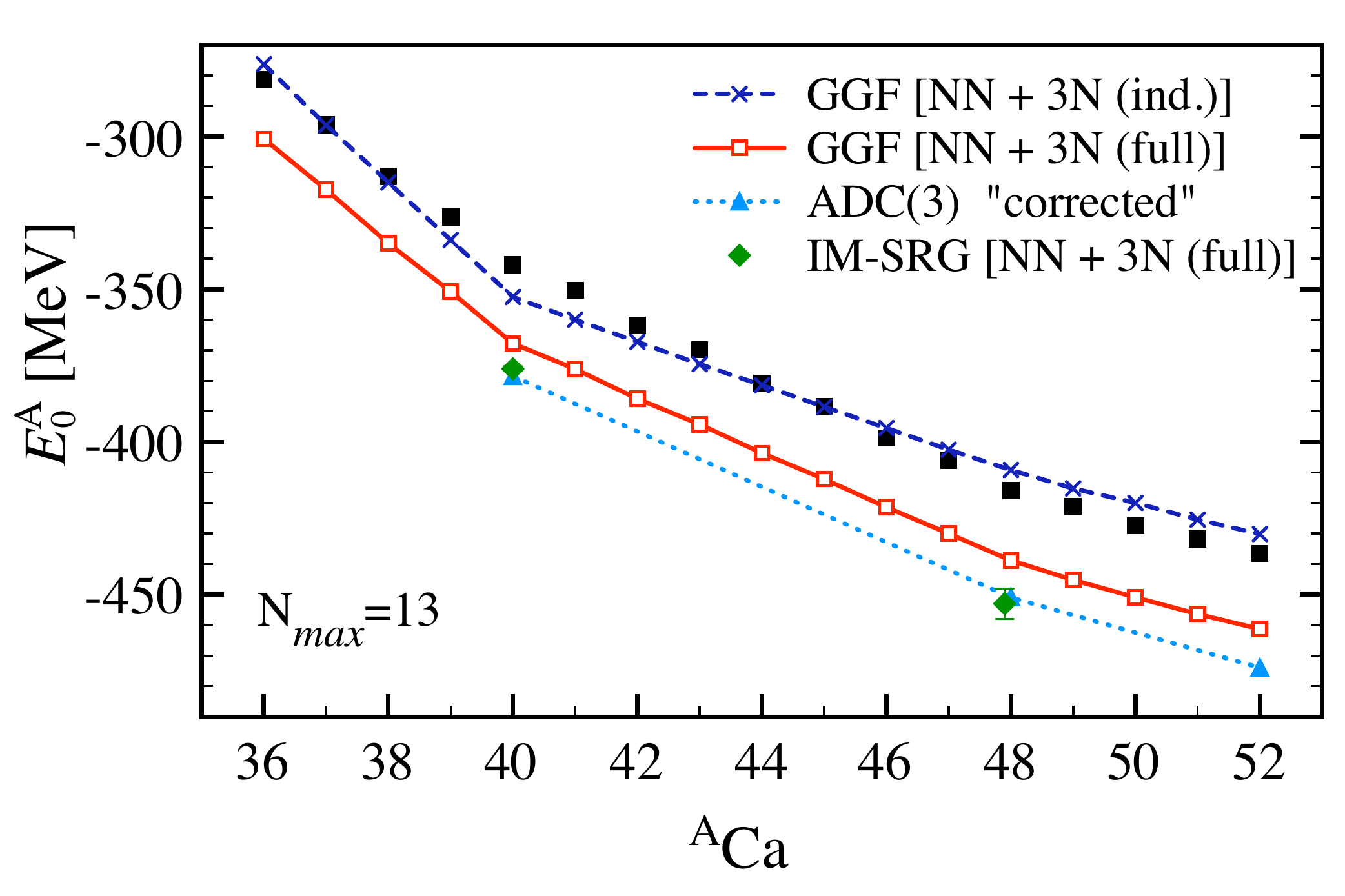}
\hspace{.7 cm}
\includegraphics[width=0.45\textwidth,height=0.56\textwidth,clip]{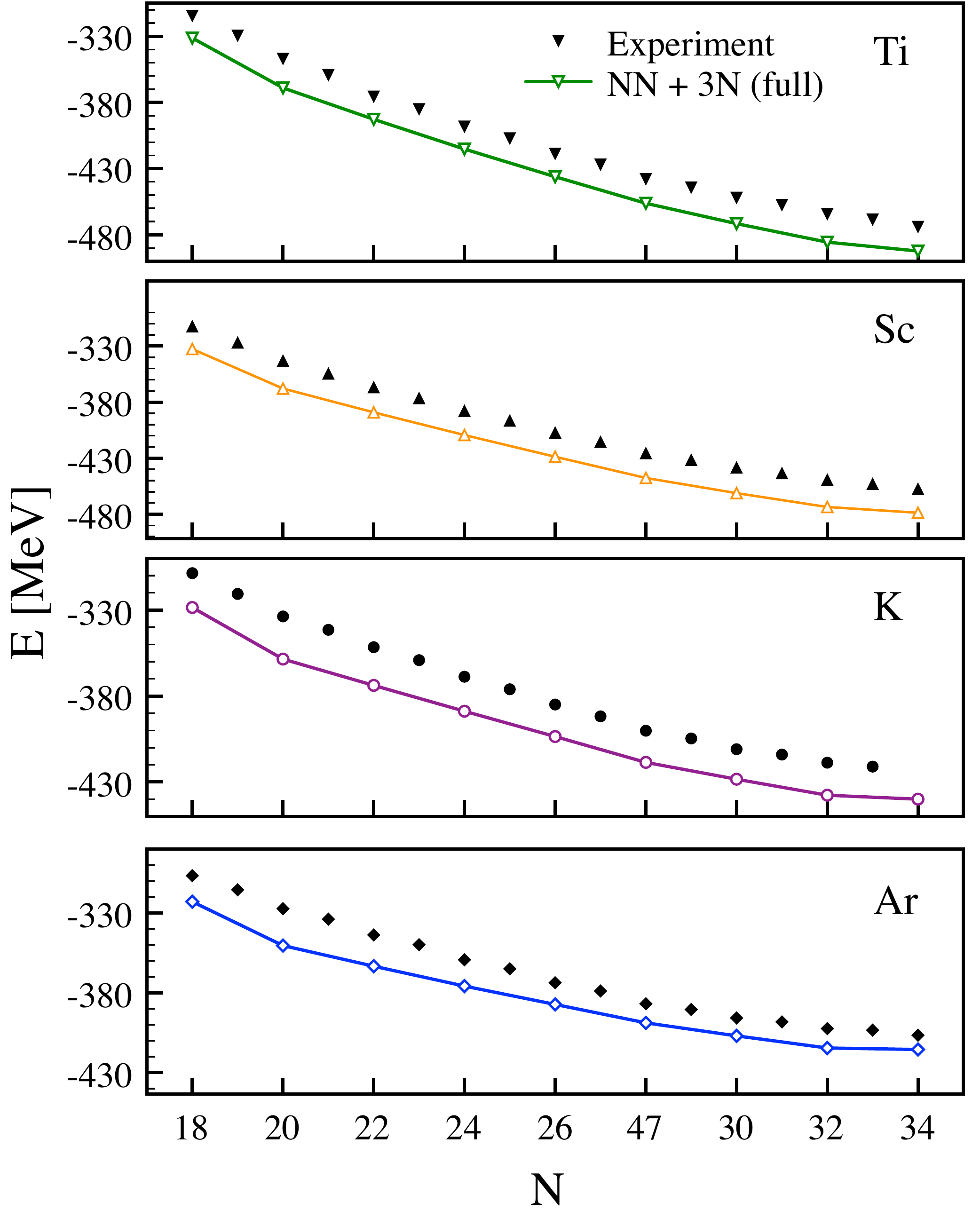}
\caption{~\cite{Soma:2013rc}. \small
{\em Left}: 
Binding energies of Ca isotopes obtained from induced and full 3NF Hamiltonians, compared to the experiment.
The triangle (with dotted light-blue line) and diamond symbols refer to Dyson-ADC(3) and IM-SRG~\cite{Hergert2013a} results for the closed shell cases. These confirm that the effects of the second order truncation in Gorkov calculations amount to a small constant shift of binding energies, which does not invalidate our general conlcusions.
{\em Right}: Predicted binding energies for Ar, K, Sc and Ti isotopic chains for the full 3NF Hamiltonian (full lines).  In all plots, experimental values are indicated by full black symbols and are taken from~\cite{AME2012tables,Gallant2012prl,Wienholtz:2013nya}. Calculations are from second order Gorkov-GF, except where indicated, and with chiral interactions evolved to $\lambda_{SRG}=2.0 \, \text{fm}^{-1}$~\cite{Soma2014s2n}.
}
\vspace{-1.5cm}
\label{fig:Ca_chains}       
\end{center}
\end{figure}

The newly introduced Gorkov-GF method allows for the first time ab-initio calculations of adjacent isotopes within large portions of the nuclear chart.
The importance of leading order chiral 3NFs in reproducing the correct trend of binding energies for the oxygen nuclides is then confirmed for heavier isotopes. This is shown in the left panel of  Fig.~\ref{fig:Ca_chains}, which compares induced and full Hamiltonians for Ca isotopes. The trend of the binding energies is predicted incorrectly by the induced 3NFs alone (blue dashed line), which generate a wrong slope and exaggerate the kink at $^{40}$Ca. This problem is fully amended by the inclusion of NNLO chiral 3NFs (red full line). However, the latter introduce additional attraction that results in a systematic over binding of ground-states throughout the whole chain. Analogous results for Ar, K, Sc and Ti  isotopic chains are shown in the left panel for full 3NFs and lead to the same conclusion regarding the role of the initial chiral 3NF: it provides the correct trend of binding energies but it generates a rather constant over binding throughout this mass region.

\begin{figure}[t!]
\begin{center}
\includegraphics[width=0.5\textwidth,clip]{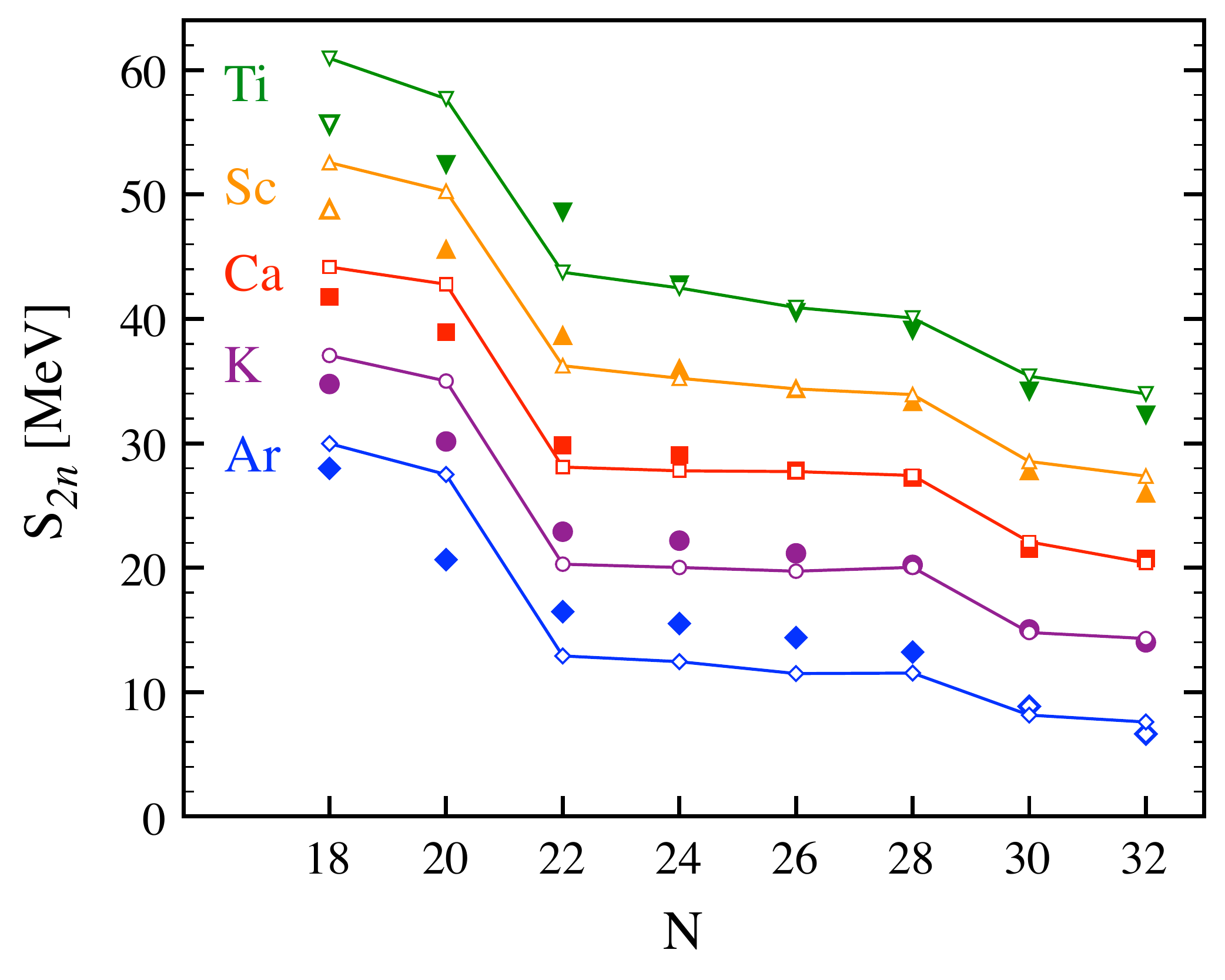}
\caption{ \small
 Two-neutron separation energies, {\em S}$_{2n}$, along Ar, K, Ca, Sc and Ti isotopic chains. The experimental values (isolated symbols)~\cite{AME2012tables,Gallant2012prl,Wienholtz:2013nya} are compared to predictions for the full 3NF Hamiltonian with a SRG cutoff of $\lambda_{SRG}=2.0 \, \text{fm}^{-1}$ (full lines). Results are for  second order Gorkov calculations and the values for K, Ca, Sc and Ti are respectively shifted by +5~MeV, 10~MeV, 15~MeV and 20~MeV for display purposes. From  Ref.~\cite{Soma2014s2n}.
}
\label{fig:S2n_all}       
\end{center}
\end{figure}

The systematic of {\em S}$_{2n}$ obtained with the NN plus full 3NF is displayed in Fig.~\ref{fig:S2n_all} along Ar, K, Ca, Sc and Ti isotopic chains, up to N=32. When the neutron chemical potential lies within the {\em pf} shell, predicted {\em S}$_{2n}$ reproduce the experiment to good accuracy. The results are quite remarkable considering that NN+3N chiral interactions have been fitted solely to few-body data up to $A=4$.
 Still, the quality slightly deteriorates as the proton chemical potential moves down into the {\em sd} shell, that is, going from Ca to K and Ar elements. The increasing underestimation of the {\em S}$_{2n}$ is consistent with a too large gap between proton {\em sd} and {\em pf} major shells, which prevents quadrupole neutron-proton correlations to switch on when the two chemical potentials sit on both sides of the gap.
This is supported by the results found for NN-N$^3$LO potential in Refs.~\cite{Barbieri2009ni,Barbieri2009prl} and in Fig.~\ref{fig:Ca44_SF}.
The too large jump of the {\em S}$_{2n}$ between N=20 and N=22 is visible for all elements and becomes particularly pronounced as one moves away from the proton magic $^{40}$Ca nucleus where the experimental jump is progressively washed out. At N=18, the situation deteriorates going from $^{38}$Ca to $^{39}$Sc and $^{40}$Ti, when the proton chemical potential moves up into the {\em pf} shell (but not going to $^{37}$K and $^{36}$Ar). This is again consistent with an exaggerated  shell gap between {\em sd} and {\em pf} shells. These  finding suggest that repulsive effects form higher chiral cutoffs or improved 3NFs may become important in heavy nuclei.

We note that the present calculations may be sensible to the truncations in three-body space performed during the SRG evolution. Nevertheless, the error introduced is expected to be small and only at large values of N for isotopes around Ca~\cite{Binder2014ccSn132}.

\section{Conclusions}  ~ 

This talk reviewed the recent developments in ab-initio calculations of finite nuclei based on the self-consistent Green's function method. The SCGF formalism offers important opportunities in advancing the theory of exotic isotopes by providing direct theoretical links between nuclear structure and reactions. Such calculations yield both binding energies and the single particle spectral function, which is probed in most nuclear structure experiments. At the same time they provide a path to generate optical potentials from first principles and at intermediate energies (up to $\approx$100~MeV), where ab-initio methods are currently lacking.

Successful ab-initio applications based on SCGF have become possible very recently following a series of technical advances~\cite{Barbieri:2007Atoms,Barbieri2009ni,Soma:2013rc,Cipollone2013prl} and the availability of chiral interactions evolved to low-momentum scales~\cite{Jurgenson2009prl,Roth2012prl}.  Of particular relevance are the recent extension to three- and many-body forces~\cite{Carbone2013tnf,Carbone2013snm,Cipollone2013prl} and the introduction of the Gorkov SCGF formalism that allows for the first time calculations of semi-magic nuclei through large portions of the nuclear chart~\cite{Soma:2011GkvI,Soma:2013rc,Soma2014GkvII}.  Applications to scattering have so far been limited to pilot studies of proton-nucleus scattering~\cite{Barbieri:2005NAscatt} and microscopic optical potentials~\cite{Waldecker:2011by}.

The SCGF  self-energies were calculated for $^{40}$Ca, $^{48}$Ca and $^{60}$Ca and  compared with potentials from the 
DOM that were obtained from fitting elastic-scattering and bound-state data. The microscopic FRPA results explain many features of the empirical DOM potentials and provides several suggestions to improve their functional, including the exploitation of parity and angular momentum dependence, which stem from the different fillings of core orbits. The non-locality of the calculated self-energies has also suggested further developments of the DOM potential~\cite{Mahzoon2014prl}. The NN tensor force has also been seen to strongly enhance the absorption in elastic scattering at energies above 20-30~MeV.

The recent ab-initio calculations of Ref.~\cite{Cipollone2013prl,Hergert2013prl,Soma2014s2n} nicely validate chiral Hamiltonians within a wide range of masses. In particular, these prove the capability of the leading order initial 3NFs (up to NNLO) to predict the trend of binding energies along full chains of open shell isotopes and at large neutron-proton asymmetry. 
To our knowledge, Ref.~\cite{Soma2014s2n} provides for the first time a study of several adjacent isotopic chains based on first principles. These results make evident that presently available interactions generate a systematic over binding of $\approx$1~MeV/A throughout the Ca region and this is further confirmed for heavier closed shell nuclides in~Ref.~\cite{Binder2014ccSn132}.
Overall, this indicates that repulsive effects form higher chiral cutoffs or from four-body interactions may become important in heavy nuclei. Improved 3NFs, at N$^3$LO and beyond, may also have an impact on improving this situation.  

To date, the ab-initio SCGF is a very promising approach which is being extended in scope. A most compelling challenge for the coming future will be to provide reliable (and converging) microscopic calculations of optical potentials, so that these can be exploited in the analysis of nuclear structure experiments.  Other possible directions can involve linking the SCGF formalism to the phenomenological shell model.
In this respect, the ab-initio calculations of effective interactions is an application of current interest~\cite{Bogner:2014baa,Jansen:2014qxa} that has already been employed in approximate form within the SCGF formalism~\cite{Barbieri2009prl}. A consistent approach can also be used to extract corresponding effective charges~\cite{BarbieriOtsukaInPrep}.

\ack ~

It is a pleasure to thank
A. Carbone, A. Cipollone,
M. Degroote,
W. H. Dickhoff, T. Duguet,
M. Hjorth-Jensen,
P. Navr\'atil,
A. Polls, A. Rios-Huguet,
V. Som\`a,
D. Van Neck and S. Waldecker
for several discussions and for all the fruitful collaborations that led to the above results.
This work was supported by the United Kingdom Science and Technology Facilities Council (STFC) under Grants No. ST/I003363/1, ST/J000051/1 and ST/L005816/1.
Calculations were performed in part using HPC resources from the DiRAC Data Analytic system at the University of Cambridge (BIS National E-infrastructure capital grant No. ST/K001590/1 and STFC grants No. ST/H008861/1, ST/H00887X/1, and ST/K00333X/1).

\section*{References} ~
 \bibliography{MBT17lib}

\end{document}